\documentclass[twocolumn,showpacs,preprintnumbers,amsmath,amssymb,prx,superscriptaddress,longbibliography]{revtex4-1}
\usepackage{bbm}
\usepackage{mathrsfs}
\usepackage{graphicx}
\usepackage{dcolumn}
\usepackage{bm}
\usepackage{amsmath}
\usepackage{amsfonts}
\usepackage{color}
\usepackage[colorlinks=true,linkcolor=magenta,urlcolor=magenta,citecolor=cyan,anchorcolor=blue]{hyperref}
\usepackage{floatrow}

\begin{document}

\title{Second-order optical response of superconductors induced by supercurrent injection}
\author{Linghao Huang}
\affiliation{State Key Laboratory of Surface Physics and Department of Physics, Fudan University, Shanghai 200433, China}
\author{Jing Wang}
\thanks{wjingphys@fudan.edu.cn}
\affiliation{State Key Laboratory of Surface Physics and Department of Physics, Fudan University, Shanghai 200433, China}
\affiliation{Institute for Nanoelectronic Devices and Quantum Computing, Fudan University, Shanghai 200433, China}
\affiliation{Zhangjiang Fudan International Innovation Center, Fudan University, Shanghai 201210, China}

\begin{abstract}
We develop a theory of the nonlinear optical responses in superconducting systems in the presence of a dc supercurrent. The optical transitions between particle-hole pair bands across the superconducting gap are allowed in clean superconductors as the inversion symmetry breaking by supercurrent. Vertex correction is included in optical conductivity to maintain the $U(1)$ gauge symmetry in the mean-field formalism, which contains the contributions from collective modes. We show two pronounced current-dependent peaks in the second-harmonic generalization $\sigma^{(2)}(2\omega,\omega,\omega)$ at the gap edge $2\hbar\omega=2\Delta$ and $\hbar\omega=2\Delta$ and one in the photocurrent effect $\sigma^{(2)}(0,\omega,-\omega)$ at $\hbar\omega=2\Delta$, all of which diverge in the clean limit. We demonstrate this in the models of a single-band superconductor with $s$-wave and $d$-wave pairings, and Dirac fermion systems with $s$-wave pairing. Our theory predicts that the current-induced peak in $\text{Im}[\sigma^{(2)}(\omega)]$ is proportional to the square of the supercurrent density in the $s$-wave single-band model, with the same order of magnitude as the recent experimental observation of second-harmonic generation in NbN by Nakamura \emph{et al.} [{\color{blue}Phys. Rev. Lett. \textbf{125}, 097004 (2020)}]. Supercurrent induced nonlinear optical spectroscopy provides a valuable toolbox to explore novel superconductors.
\end{abstract}

\date{\today}

\maketitle

\section{Introduction}
\label{introduction}

The linear and nonlinear optical spectroscopy methods are well established as a powerful experimental technique to explore quantum materials, in particular superconductors~\cite{orenstein2012,basov2011,tinkham1974,basov2005}, which have long been the focus of research. In semiconductors, optical spectroscopy provides direct and nondestructive measurements of pure spin currents~\cite{wang2010,werake2010,wang2008}. In the context of superconductors, the linear optical measurements can unveil the characteristic nature of the superconducting state, for example, by probing spectral weight transfer~\cite{molegraaf2002} and measuring the superconducting gap size~\cite{degiorgi1994,pronin1998}. Theoretically, the optical properties can be characterized by optical conductivity $\sigma(\omega)$, which can be calculated from microscopic considerations. The earliest analysis of the linear optical response for superconductors was made by Mattis and Bardeen in 1958~\cite{mattis1958}, who stated that in a single-band Bardeen-Cooper-Schrieffer (BCS) superconductor, the optical absorption (namely the real part of $\sigma^{(1)}(\omega)$) is absent when the photon energy of incident light lies at or below the band gap in the dirty limit (superconducting coherence length $\xi_0$ $\gg$ mean free path $\ell$). Furthermore, optical transitions even vanish at any finite frequency without the mediation of impurity~\cite{mahan2000}, owing to the band structure of the normal state: if the dispersion of the electron satisfies $\epsilon_{\mathbf{k}}=\epsilon_{-\mathbf{k}}$, which can be guaranteed by inversion symmetry $\mathcal{I}$ or time-reversal symmetry $\mathcal{T}$, when a Cooper pair is broken by light, the two electrons will have the same velocity but in opposite directions, resulting in zero net current. The Mattis–Bardeen theory, together with its generalization to arbitrary $\ell$~\cite{zimmermann1991}, has been tested very successfully in explaining the linear optical properties of many superconductors. Recently, the optical transitions in a multiband superconductor with $\mathcal{I}$ symmetry were theoretically shown to be allowed, leading to a contribution to optical conductivity in the clean limit ($\xi_0\ll\ell$)~\cite{ahn2021}.

Meanwhile, the nonlinear response of superconductors has attracted considerable interest. The second-order optical effects exist in superconductors with intrinsic broken $\mathcal{I}$ symmetry of the band structure or superconducting pairing order parameter~\cite{xu2019,watanabe2022a,tanaka2023}. The second-order nonlinear conductivity shows divergent behavior in the low-frequency limit, which is analogous to the linear order case and unique to superconductors. Apart from quasiparticles, the excitation of collective modes in superconductors could also contribute to the optical response~\cite{cea2016a,tsuji2016,matsunaga2017,cea2018,yang2018,yang2019}, where a well-known example is the third-harmonic generation of Higgs mode~\cite{matsunaga2014}. As a scalar excitation, the Higgs mode can couple to a gauge field at least in second order, giving rise to a photocurrent with triple frequency of incident light. Therefore, the third-harmonic generation of superconductors would originate from both quasi-particle and Higgs mode excitation.

Recently, there is growing interest in the effect of a supercurrent on the optical response of superconductors. With the help of a dc supercurrent, the Higgs mode can be linearly excited by an electromagnetic field~\cite{moor2017,puviani2020a}, leading to the linear optical response. Meanwhile, a dc supercurrent breaks $\mathcal{I}$ symmetry extrinsically, and thus makes $\epsilon_{\mathbf{k}}\neq \epsilon_{-\mathbf{k}}$, enabling the nonzero optical response of clean single-band superconductors~\cite{crowley2022,papaj2022}. Experimentally, an optical response induced by a supercurrent was measured in $\text{Nb}_3\text{Sn}$~\cite{yang2019a,vaswani2020} and NbN~\cite{nakamura2019,nakamura2020} film, where the enhanced optical response near the gap edge by a supercurrent was observed in the linear order optical effect, and the response peaks of second-harmonic generation (SHG) was detected.

It has been a longstanding proposal that gauge invariance should be maintained when calculating optical responses~\cite{bardeen1957}. In the BCS mean-field formalism, the $U(1)$ gauge symmetry is broken, causing local charge nonconservation of the optical response kernel and an unphysical result in the longitudinal direction~\cite{schrieffer1999}. Nambu extended the vertex correction method widely used in quantum electrodynamics to resolve this issue~\cite{nambu1960}, whose approach has been adopted in recent research on the linear optical response of a superconductor, and the vertex correction dramatically changes the linear optical conductivity~\cite{dai2017,papaj2022}. All of these previous studies focus on the linear optical response; a natural question arises as to how gauge invariance affects the nonlinear optical response of superconductors, which is the main theme of the current paper. We develop a theory of the nonlinear optical responses in clean superconducting systems in the presence of a dc supercurrent, with the vertex correction reflecting electron-electron interaction included. We show that the second-order optical conductivity $\sigma^{(2)}(\omega)$ depends on the nature of the normal state as well as the type of superconducting pairing, which is demonstrated in the model of single-band superconductor with $s$-wave and $d$-wave pairings, and Dirac fermion systems with $s$-wave pairing. Remarkably, we find that the peak value of $\sigma^{(2)}(\omega)$ in an $s$-wave superconductor has the same order of magnitude as the SHG experimental measurement~\cite{nakamura2020}, which means the proposed effect here may very well already have been observed. The supercurrent flow may be induced by an external magnetic field through the Meissner effect, and thus the nonlinear optical spectroscopy provides a valuable toolbox to explore the superconducting state. 

The paper is organized as follows. Section~\ref{diagrammatic} briefly reviews the diagrammatic method applied to the superconductor in mean-field formalism. Section~\ref{second-order} presents the formalism of the gauge-invariant second-order nonlinear optical response in superconductors by including vertex correction, which can be readily applied to higher-order responses. In Section~\ref{model}, we calculate second-order nonlinear optical conductivity in several different superconductor models, namely, the lattice model of a single-band superconductor with $s$-wave and $d$-wave pairings, and Dirac fermion systems with proximity effect from an $s$-wave superconductor. Finally, we conclude our work with some further discussions in Section~\ref{discussion}. Some auxiliary materials are relegated to the Appendices.

\section{Diagrammatic method applied to superconductor}
\label{diagrammatic}

\begin{figure}[t]
\begin{center}
\includegraphics[width=3.4in,clip=true]{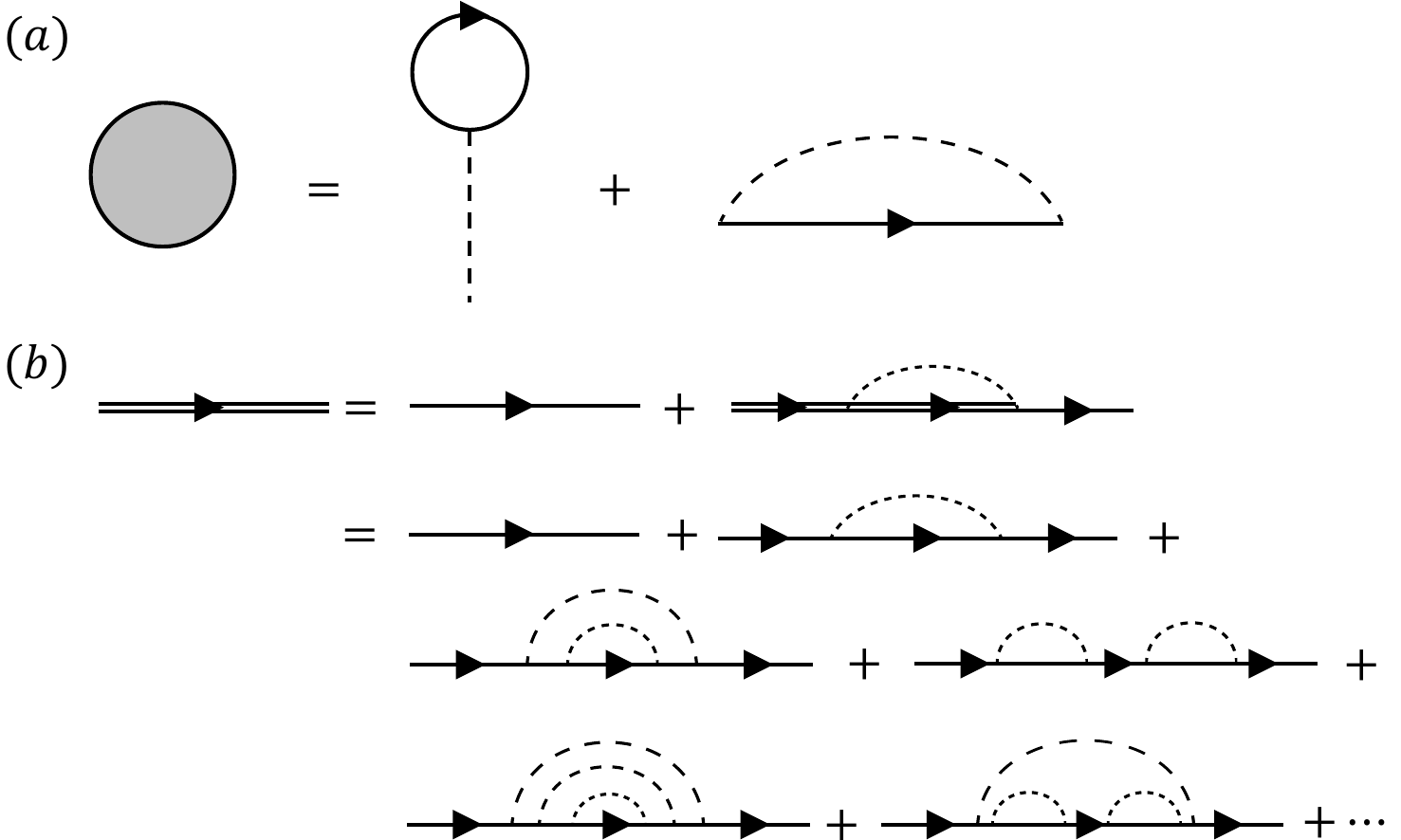}
\end{center}
\caption{(a) Feynman diagram of self-energy correction for the electron Green's function up to first order. The first, on the right -hand side, represents the Hartree term, and the second denotes the Fock term. The single solid line represents the electron Green's function; the dashed line represents the electron-electron interaction. (b) Feynman diagram of the Green's function corrected by interaction, which is represented by a double solid line. Only off-diagonal correction is considered, and there is no crossing of the interaction line.}
\label{fig1}
\end{figure}

We start by briefly reviewing the mean-field theory of a superconductor, specifically the diagrammatic method of the mean-field gap equation. The BCS mean-field theory can be understood as the self-consistent Hartree-Fock (SCHF) approximation~\cite{nambu1960,bogolyubov1959}. To see this explicitly, we consider a two-dimensional system with an effective attracting two-body interaction between the electrons, while the extension to three-dimensional case is straightforward. The interacting Hamiltonian is $H=\sum_{\mathbf{k},\sigma} \epsilon_k c_{\mathbf{k},\sigma}^\dagger c_{\mathbf{k},\sigma} - \sum_{\mathbf{k},\mathbf{k'}} V_{\mathbf{k},\mathbf{k'}} c_{\mathbf{k},\uparrow}^\dagger c_{-\mathbf{k},\downarrow}^\dagger c_{-\mathbf{k'},\downarrow} c_{\mathbf{k'},\uparrow} $.
In the Nambu basis, the interaction term can be written as $ \sum_{\mathbf{k,k'}} V_{\mathbf{k},\mathbf{k'}} [\Psi_{\mathbf{k}}^\dagger \tau_3 \Psi_{\mathbf{k'}}] [\Psi_{\mathbf{k'}}^\dagger \tau_3 \Psi_{\mathbf{k}}] $,
where $ \Psi_{\mathbf{k}}= (c_{\mathbf{k},\uparrow}, c^\dagger_{\mathbf{-k},\downarrow})^T $ is the Nambu spinor operator and $\tau_3$ is the Pauli matrix. We approximate the four-fermion interaction by a quadratic term. Then, up to first order of the interaction, the self-energy of the electron can be written as,
\begin{eqnarray}\label{selfene}
    \Sigma(k_0,\mathbf{k}) &=& \Sigma_\text{H}+\Sigma_\text{F}(k_0,\mathbf{k}), 
    \\
    \Sigma_\text{H} &=& -\frac{1}{\beta} \tau_3 \text{Tr}\left[\sum_{k'_0} \int \frac{\mathrm{d}^2 \mathbf{k}'}{(2\pi)^2} V_{\mathbf{k}',\mathbf{k}'} G(k'_0,\mathbf{k}') \tau_3\right], 
    \nonumber
    \\
    \Sigma_\text{F}(k_0,\mathbf{k}) &=& \frac{1}{\beta}\sum_{k'_0} \int \frac{\mathrm{d}^2\mathbf{k}'}{(2\pi)^2} V_{\mathbf{k}-\mathbf{k}',\mathbf{k}} \tau_3G(k_0-k'_0,\mathbf{k}-\mathbf{k}') \tau_3,
    \nonumber
\end{eqnarray}
where $G(k_0,\mathbf{k}) = [i k_0 \tau_0 - H_0(\mathbf{k}) - \Sigma(k_0,\mathbf{k})]^{-1}$ is the Matsubara Green's function containing self-energy correction in the Nambu basis, $k_0$ is the temporal component of the momentum, $H_0(\mathbf{k}) = \epsilon_k \tau_3$ is the free Hamiltonian of the original band, and we have implied that $\epsilon_{\mathbf{k}}=\epsilon_{-\mathbf{k}}$. The corresponding Feynman diagrams of the Hartree $\Sigma_\text{H}$ and Fock $\Sigma_\text{F}(k_0,\mathbf{k})$ terms are shown in Fig.~\ref{fig1}(a). Since the diagonal correction of self-energy only modifies the band structure and dose not contribute to pairing potential, we ignore this part, and define $\Sigma(k)=\Delta(k)\tau_1$. On this account, the Feynman diagram of the corrected Green's function is constructed by adding an interaction line that can transfer finite momentum; see Fig.~\ref{fig1}(b). A typical character of this diagram is that no crossing of the interaction line is considered. Therefore, we retrieve the Bogoliubov-de Gennes (BdG) mean-field Hamiltonian from SCHF approximation as
\begin{equation}\label{BdG1}
    H_{\text{BdG}}=\sum_{\mathbf{k}}\Psi_{\mathbf{k}}^\dagger 
    \begin{pmatrix}
    \epsilon_{\mathbf{k}} & \Delta(\mathbf{k}) \\
    \Delta(\mathbf{k}) & -\epsilon_{-\mathbf{k}}
    \end{pmatrix}
    \Psi_{\mathbf{k}}.
\end{equation}
At zero temperature, the corresponding self-consistent gap equation is obtained by integrating out $k'_0$,
\begin{equation}\label{gapeqn}
    \Delta(\mathbf{k}) =  \int \frac{\mathrm{d}^2\mathbf{k'}}{(2\pi)^2} \frac{V_{\mathbf{k},\mathbf{k'}} \Delta(\mathbf{k'})}{2\sqrt{\epsilon^2_\mathbf{k'}}+\Delta(\mathbf{k'})^2}.
\end{equation}
If $ V_{\mathbf{k},\mathbf{k'}} $ can be approximate to be constant near the Fermi surface, $ \Sigma(\mathbf{k}) $ will become a constant matrix, which is the case for the $s$-wave BCS superconductor. If $ V_{\mathbf{k},\mathbf{k'}} $ can be factorized as $ V \phi_\mathbf{k}\phi_\mathbf{k'} $, and $\phi_\mathbf{k}$ takes the form $(\cos{k_x}-\cos{k_y})$, $\Sigma(\mathbf{k})$ turns out to be a constant matrix multiplied by the factor $\phi_\mathbf{k}$, which is the $d$-wave case~\cite{coleman2015}.

When the superconductor carries a dc supercurrent, which may be from an external source or induced by a magnetic field, a Cooper pair will get a nonzero total momentum $2\mathbf{q}$. Then we change the Nambu spinor to be $\Psi_{\mathbf{k}}=(c_{\mathbf{k}+\mathbf{q},\uparrow}, c^\dag_{\mathbf{-k}+\mathbf{q},\downarrow})^T$, and the BCS mean-field Hamiltonian will become~\cite{dai2017},
\begin{equation}\label{BdG2}
    H_{\text{BdG}}=\sum_{\mathbf{k}}\Psi_{\mathbf{k}}^\dagger 
    \begin{pmatrix}
    \epsilon_{\mathbf{k}+\mathbf{q}} & \Delta(\mathbf{k}) \\
    \Delta(\mathbf{k}) & -\epsilon_{-\mathbf{k}+\mathbf{q}}
    \end{pmatrix}
    \Psi_{\mathbf{k}}.
\end{equation}
The eigenvalues of the BdG Hamiltonian are
\begin{equation}\label{eigen2}
     E_{\mathbf{k},\pm}=\frac{\epsilon_{\mathbf{k}+\mathbf{q}}-\epsilon_{-\mathbf{k}+\mathbf{q}}}{2} \pm \sqrt{\left(\frac{\epsilon_{\mathbf{k}+\mathbf{q}}+\epsilon_{-\mathbf{k}+\mathbf{q}}}{2}\right)^2+\Delta(\mathbf{k})^2}.
\end{equation}
The gap equation is almost the same as Eq.~(\ref{gapeqn}), except that $ \epsilon_\mathbf{k} $ should be replaced by $\left(\epsilon_{\mathbf{k}+\mathbf{q}}+\epsilon_{-\mathbf{k}+\mathbf{q}}\right)/2$. Without loss of generality, we assume the supercurrent is along the $x$ axis, namely $\mathbf{q}=q\hat{\mathbf{x}}$ hereafter.

\section{Vertex correction and gauge-invariant second-order optical response}
\label{second-order}

\begin{figure}[t]
    \begin{center}
    \includegraphics[width=3.3in,clip=true]{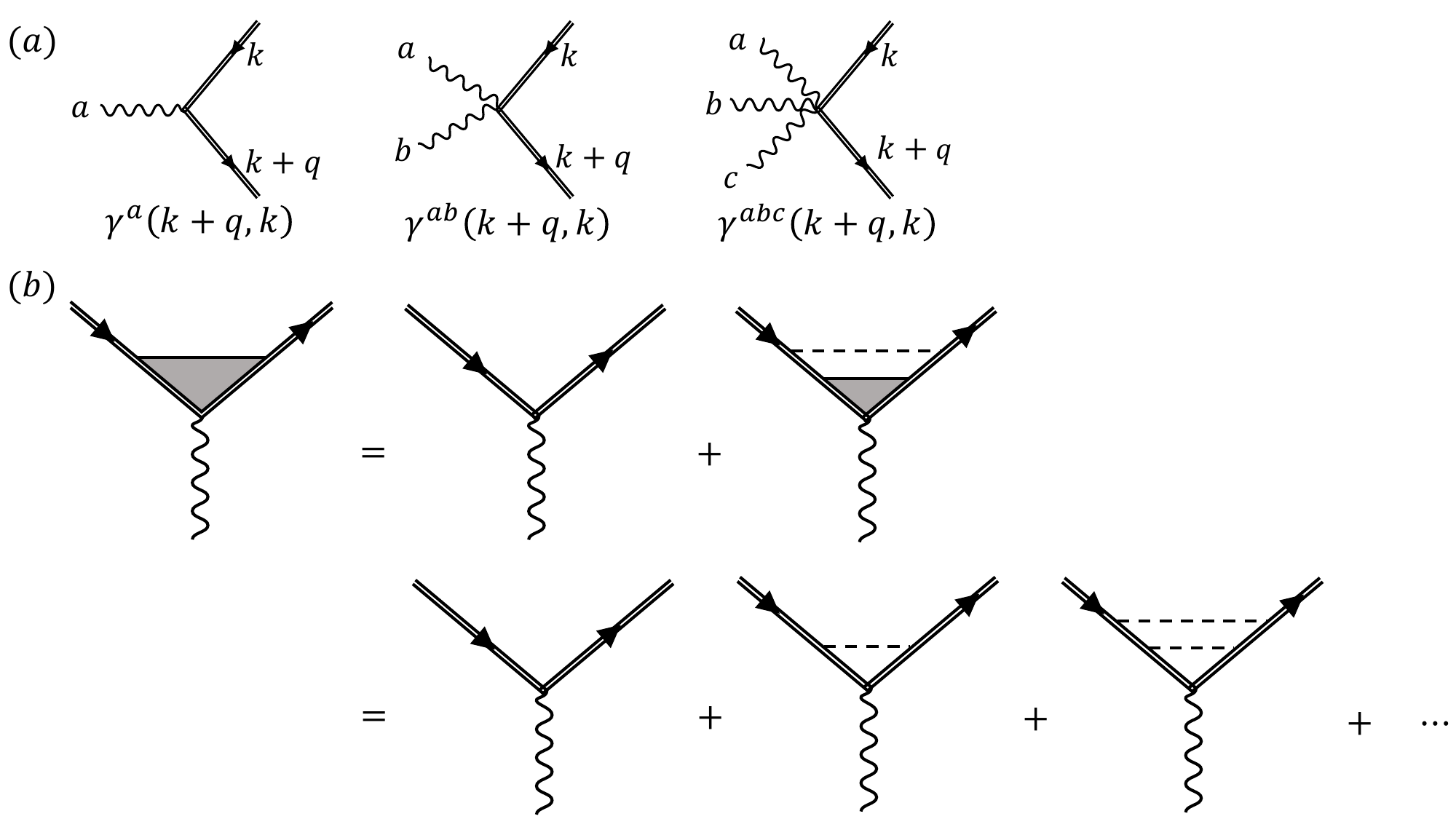}
    \end{center}
    \caption{(a) Feynman diagrams for bare vertex functions in calculating the first and second optical responses. The wavy line represents the electromagnetic field. (b) Feynman diagram for self-consistent equation of vertex correction under no-crossing approximation. These types of diagrams are also called ladder diagrams.}
    \label{fig2}
\end{figure}

Now we are ready to study the nonlinear optical response of superconductors. In the velocity gauge, an electromagnetic field couples to a superconductor through minimal coupling: $\mathbf{k} \rightarrow \mathbf{k}-q'\mathbf{A}/\hbar$ in the diagonal sector of BdG Hamiltonian, where $q'=-e$ $(e)$ is the electron (hole) charge. We focus on the limit of a weak, monochromatic and uniform external electromagnetic field. The generalized velocity operators are defined by~\cite{watanabe2022a},
\begin{eqnarray}\label{gamma}
    \gamma_{a_1 a_2 \cdots a_n} & = & \left(-\frac{1}{e}\right)^n \left.\frac{\partial^n H_{\text{BdG}}(\mathbf{A})}{\partial A^{a_1}\partial A^{a_2}\cdots \partial A^{a_n}}\right|_{\mathbf{A}\rightarrow 0}\\
    & = & \left(-\frac{\tau_3}{\hbar}\right)^n
    \begin{pmatrix}
        \partial_{a_1 a_2 \cdots a_n} \epsilon_{\mathbf{k}+\mathbf{q}} & 0 \\
        0 & -\partial_{a_1 a_2 \cdots a_n} \epsilon_{-\mathbf{k}+\mathbf{q}}
    \end{pmatrix},
    \nonumber
\end{eqnarray}
where $\partial_{a_1 a_2 \cdots a_n}$ is the abbreviation for $ \frac{\partial}{\partial k^{a_1}} \frac{\partial}{\partial k^{a_2}} \cdots \frac{\partial}{\partial k^{a_n}} $. These operators are represented by bare electron-photon vertex functions in the diagrammatic method, some examples of which are shown in Fig.~\ref{fig2}(a).

As mentioned previously, the gauge invariance should be maintained when calculating the linear optical response in BCS mean-field theory~\cite{papaj2022,dai2017,oh2023}, and thus the Ward-Takahashi identity needs to be satisfied. This requirement must still be fulfilled if one calculate higher order optical responses~\cite{rostami2021gauge}. This can be done by introducing a corrected electron-photon vertex~\cite{nambu1960}. In Appendix~\ref{app A}, we derive a series of Ward identities needed for the nonlinear optical response at arbitrary order. In the SCHF approximation to the first order, the electron mean-field Green's function is corrected by interaction under the no-crossing approximation. Therefore, the electron-photon vertex should also be corrected by the no-crossing interaction line due to Ward identity; see Fig.~\ref{fig2}(b). The corrected vertex function is determined by the self-consistent equation
\begin{equation}\label{Gamma}
    \begin{aligned}
     & \Gamma(p+q,p) = \gamma(p+q,p)+ \frac{1}{\beta}\sum_{k_0} \int \frac{\mathrm{d^2} \mathbf{k}}{(2\pi)^2} \\
    & [V_{\mathbf{k},\mathbf{p}} \tau_3 G(k_0+q_0,\mathbf{k}+\mathbf{q}) \Gamma(k+q,k) G(k_0,\mathbf{k})\tau_3],
    \end{aligned}   
\end{equation}
where $\gamma(p+q,p)$ and $\Gamma(p+q,p)$ are the bare and corrected vertex. For an $s$-wave superconductor, $V_{\mathbf{k},\mathbf{p}}$ is a constant, the integrand, and thus $\Gamma(p+q,p)$ in the above equation is independent of $p$, while $\Gamma(p+q,p)$ is generically determined both by $p$ and $q$ (e.g., $d$-wave case; see Appendix~\ref{app B}). In general, the corrected vertex function will contain all $\tau_1$, $\tau_2$ and $\tau_3$ components, which characterize the fluctuations of the pairing order parameter and charge density. Thus, after considering the vertex correction, the final electromagnetic response already contains the contributions from the collective modes, which is the necessary requirement for a gauge-invariant theory.

\begin{figure}[b]
\begin{center}
\includegraphics[width=3.4in,clip=true]{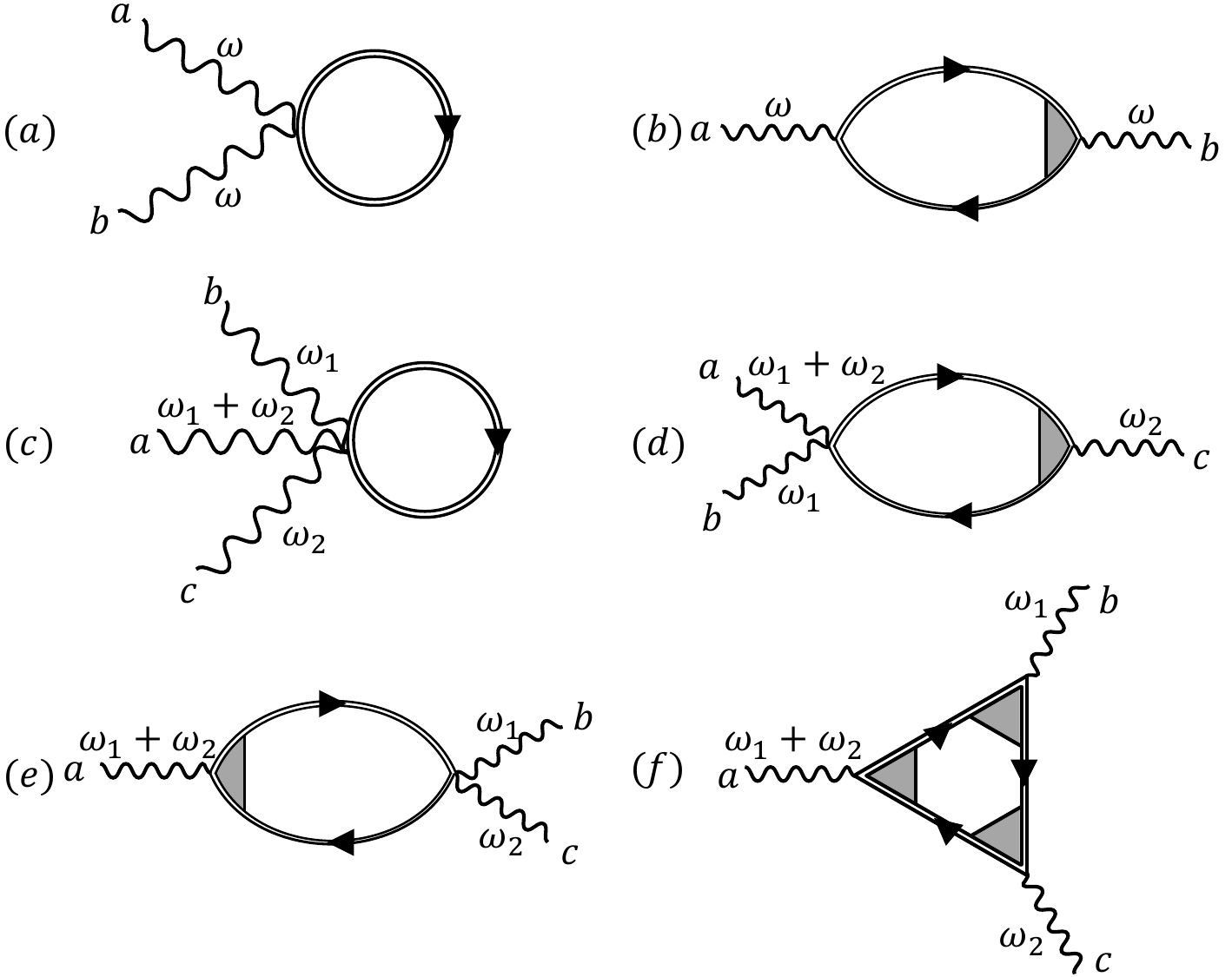}
\end{center}
\caption{(a),(b) Feynman diagrams contribute to first-order optical conductivity of the superconductor. (c)-(f) Feynman diagrams contribute to second-order nonlinear optical conductivity of the superconductor. Vertex correction under no-crossing approximation is considered. Note that to avoid double counting, there is no vertex correction in (a) and (c), and only one of the vertices is corrected in (b), (d), and (e).}
\label{fig3}
\end{figure}

The nonlinear optical responses of the superconductors can be calculated via the diagrammatic approach~\cite{parker2019,michishita2021,rostami2021a,rostami2017theory}. Here we mainly consider the clean limit $\ell\gg\xi_0$, which means the momentum of the electron remains unchanged during optical transition under uniform and monophonic light illumination. Moreover, we calculate the optical response up to second order; this method also works for a higher-order response. The first- and second-order optical conductivity are defined as
\begin{equation}\label{j}
    \begin{aligned}
    j^{(1)}_a & = \sigma^{(1)}_{ab}(\omega)E_b(\omega),    \\
    j^{(2)}_a & = \sigma^{(2)}_{abc}(\omega_1+\omega_2,\omega_1,\omega_2)E_b(\omega_1)E_c(\omega_2).    
    \end{aligned}
\end{equation}
The subscripts $a, b, c$ are $x$ or $y$ in the two-dimensional system. After vertex correction, the gauge-invariant formulas for $\sigma^{(1)}_{ab}(\omega)$ and $\sigma^{(2)}_{abc}(\omega_1+\omega_2,\omega_1,\omega_2)$ are given through Feynman diagrams shown in Fig.~\ref{fig3},
\begin{widetext}
\begin{equation}\label{sigma1}
    \sigma^{(1)}_{ab}(\omega) = \frac{i e^2}{\hbar \omega}\frac{1}{\beta}\text{Tr} \sum_{k_0} \int \frac{\mathrm{d}^2 \mathbf{k}}{(2\pi)^2} \left[ \gamma_{ab} G(k_0,\mathbf{k})+\gamma_a G(k_0+\omega_2,\mathbf{k}) \Gamma_b G(k_0,\mathbf{k}) \right],
\end{equation}
\begin{equation}\label{sigma2}
    \begin{aligned}
        & \sigma_{abc}^{(2)}(\omega_1+\omega_2,\omega_1,\omega_2)= \frac{e^3}{\hbar \omega_1 \omega_2}\frac{1}{\beta}\text{Tr} \sum_{k_0} \int \frac{\mathrm{d}^2 \mathbf{k}}{(2\pi)^2} \biggl \{\frac{1}{2} \gamma_{abc} G(k_0,\mathbf{k})+ \gamma_{ab} G(k_0+\omega_2,\mathbf{k}) \Gamma_c G(k_0,\mathbf{k}) \\
        & +\frac{1}{2}\Gamma_a G(k_0+\omega_1+\omega_2,\mathbf{k}) \gamma_{bc} G(k_0,\mathbf{k}) +\Gamma_a G(k_0+\omega_1+\omega_2,\mathbf{k}) \Gamma_b G(k_0+\omega_2,\mathbf{k}) \Gamma_c G(k_0,\mathbf{k}) +(b,\omega_1)\leftrightarrow (c,\omega_2)\biggr \} .
    \end{aligned}
\end{equation}
\end{widetext}

Note that only the temporal component of photon momentum $q$ is considered since spatial components $\mathbf{q}$ are negligibly small in the terahertz region, whose corresponding energy is comparable to the superconducting gap. The analytic continuation should be done in the Green's function as $\omega\rightarrow\omega+i\eta$, where $\eta$ is a infinitesimal positive parameter. In addition, we consider the low-temperature limit, i.e., the summation of the Matsubara frequency can be replaced by an integral. We study two processes in the second-order response: SHG and the photocurrent effect (PC)~\cite{rostami2018nonlinear}, characterized by $\sigma^{(2)}_{abc}(2\omega,\omega,\omega)$ and $\sigma^{(2)}_{abc}(0,\omega,-\omega)$, respectively. We will take both into account and show the frequency dependence of all the allowed their components in the subsequent superconductor models.

\section{Results for superconductor models}
\label{model}
 
Now we calculate second-order nonlinear optical conductivity $\sigma^{(2)}(\omega)$ in several different models of superconductors, which could exemplify the different aspects of supercurrent-enabled optical conductivity.

\subsection{$s$-wave single-band superconductor}
\label{s-wave}
 
We first consider a single-band spin-degenerate $s$-wave superconductor. A generic tight-binding model is on a square lattice with unit lattice constant, the nearest neighbor hopping $t$, the chemical potential $\mu$, and a superconducting gap $\Delta$,
\begin{equation}\label{tightbinding}
    H^{s}_{\text{TB}}(\mathbf{k})=
    \begin{pmatrix}
        \epsilon_{\mathbf{k}+q\hat{x}} & \Delta \\
        \Delta & -\epsilon_{-\mathbf{k}+q\hat{x}}
    \end{pmatrix},
\end{equation}
where $\epsilon_{\mathbf{k}}=t(2-\cos k_x-\cos k_y)-\mu$. The model has $C_{2x}$ symmetry when the supercurrent flows along the $x$ direction. The symmetry allows two nonzero independent components of $\sigma^{(1)}: \sigma^{(1)}_{xx}, \sigma^{(1)}_{yy}$, and four possible nonzero independent components of $\sigma^{(2)}: \sigma^{(2)}_{xxx}, \sigma^{(2)}_{xyy}, \sigma^{(2)}_{yxy}, \sigma^{(2)}_{yyx} $. All of the remaining components are zero. The intrinsic permutation symmetry leads to $\sigma^{(2)}_{abc}(\omega_1+\omega_2,\omega_1,\omega_2)=\sigma^{(2)}_{acb}(\omega_1+\omega_2,\omega_2,\omega_1)$, and the reality condition requires $\sigma^{(2)}_{abc}(\omega_1+\omega_2,\omega_1,\omega_2)=\left(\sigma^{(2)}_{abc}\right)^*(-\omega_1-\omega_2,-\omega_1,-\omega_2)$. Thus we obtain the following relations of nonlinear optical responses in the low-frequency regime
\begin{equation}\label{symcon}
    \begin{aligned}
        \sigma^{(2)}_{abc}(2\omega,\omega,\omega)&=\sigma^{(2)}_{acb}(2\omega,\omega,\omega), \\
        \sigma^{(2)}_{abc}(0,\omega,-\omega)&=\left(\sigma^{(2)}_{acb}\right)^*(0,\omega,-\omega).
    \end{aligned}
\end{equation}
Finally, only $\sigma^{(2)}_{xxx},\sigma^{(2)}_{xyy},\sigma^{(2)}_{yyx}$ are independent. In addition, $\sigma_{xxx}(0,\omega,-\omega)$ and $\sigma_{xyy}(0,\omega,-\omega)$ are real.

\begin{figure}[t]
\begin{center}
\includegraphics[width=3.4in,clip=true]{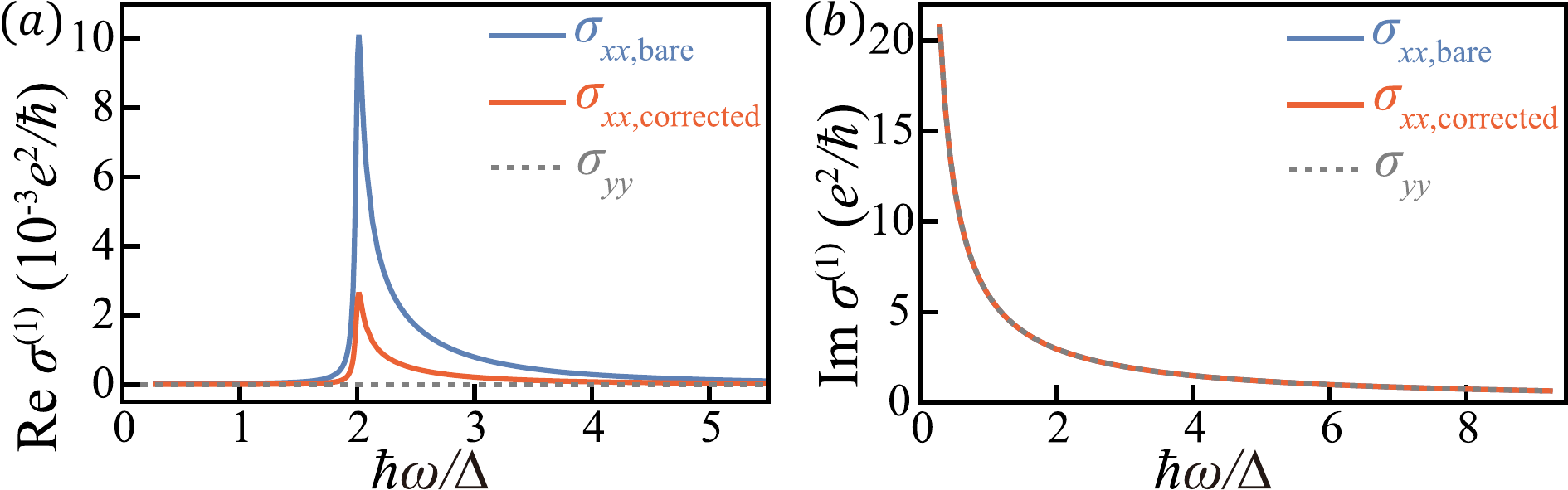}
\end{center}
\caption{$\sigma^{(1)}(\omega)$ of $s$-wave single-band model. (a) The real part of $\sigma^{(1)}_{xx}(\omega)$ without (blue solid line) and with (red solid line) vertex correction. (b) $\text{Im}[\sigma^{(1)}_{xx}(\omega)]$ without and with vertex correction. $\text{Im}[\sigma^{(1)}_{yy}(\omega)]$ (gray dashed line) is also shown.}
\label{fig4}
\end{figure}

\begin{figure}[b]
\begin{center}
\includegraphics[width=3.4in,clip=true]{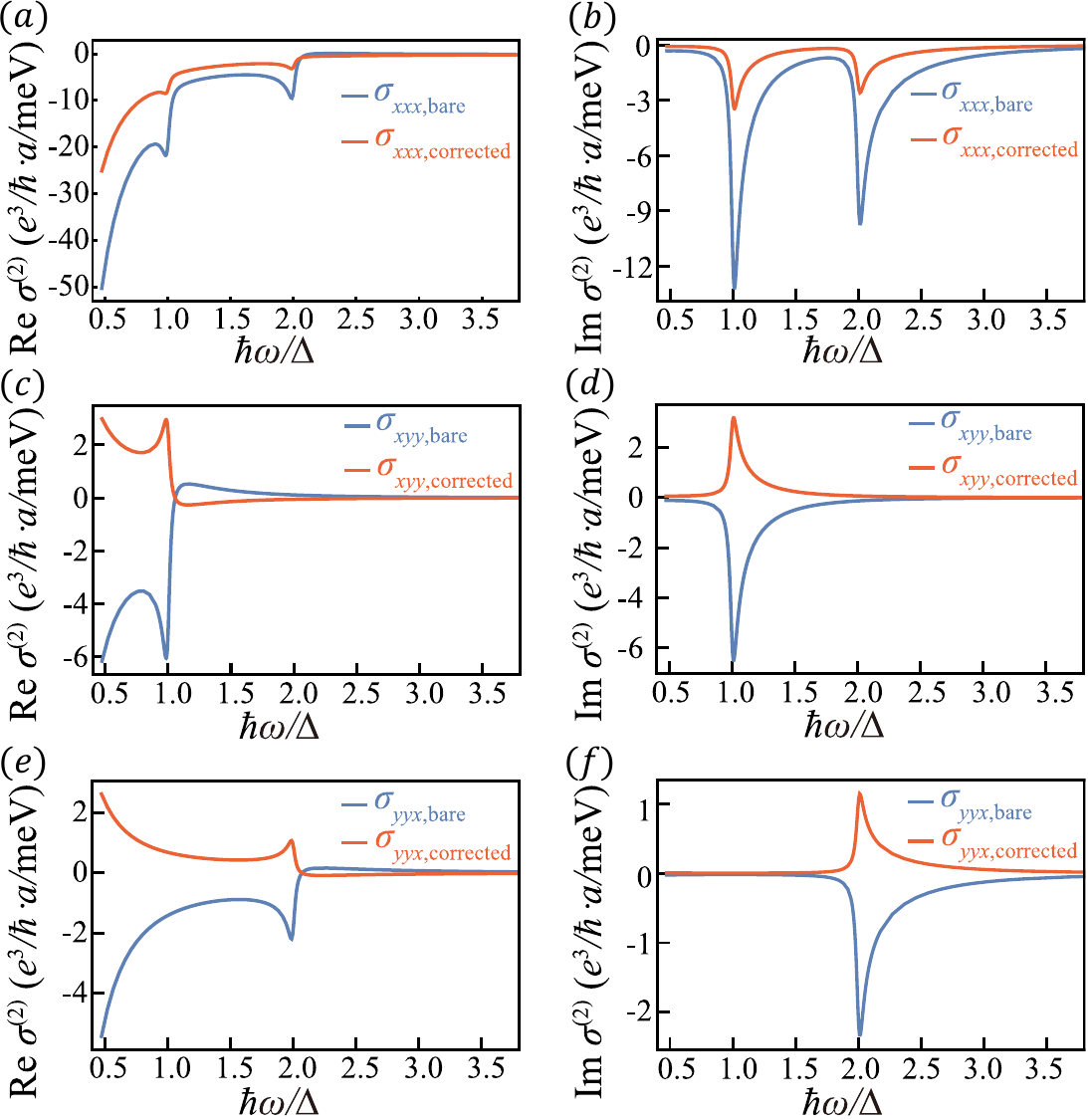}
\end{center}
\caption{$\sigma^{(2)}_{\text{SHG}}$ of the $s$-wave tight-binding model. Blue and red lines show SHG without and with vertex correction, respectively. (a),(c),(e) real part $\text{Re}[\sigma^{(2)}_{\text{SHG}}]$; (b),(d),(f) imaginary part $\text{Im}[\sigma^{(2)}_{\text{SHG}}]$.}
\label{fig5}
\end{figure}

\begin{figure}[b]
\begin{center}
\includegraphics[width=3.4in,clip=true]{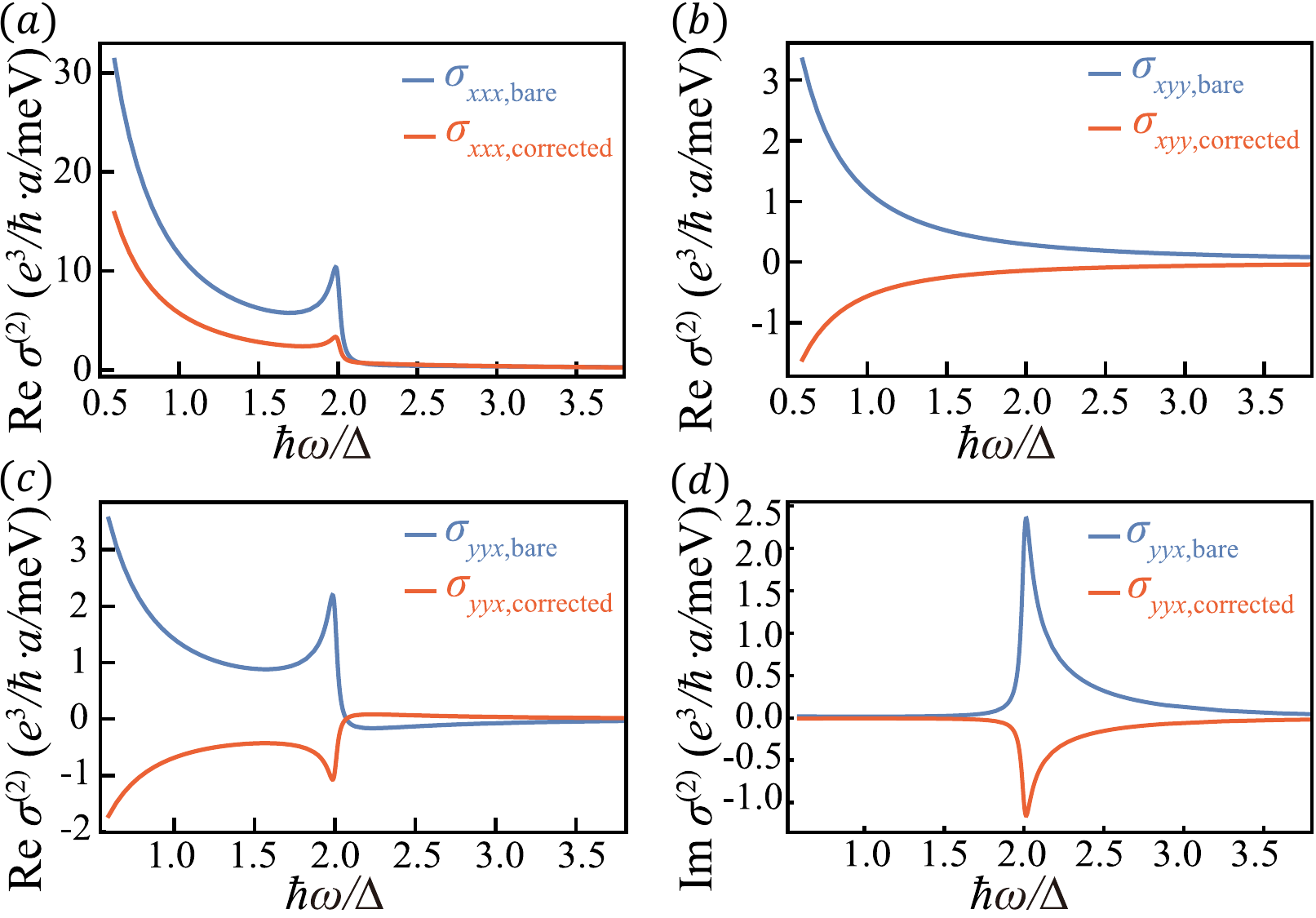}
\end{center}
\caption{$\sigma^{(2)}_{\text{PC}}$ of the $s$-wave tight-binding model. Blue and red lines show PC without and with vertex correction, respectively. (a) $\text{Re}[\sigma^{(2)}_{\text{PC},xxx}]$, (b) $\text{Re}[\sigma^{(2)}_{\text{PC},xyy}]$; both of their imaginary parts are zero. (c) $\text{Re}[\sigma^{(2)}_{\text{PC},yyx}]$. (d) $\text{Im}[\sigma^{(2)}_{\text{PC},yyx}]$.}
\label{fig6}
\end{figure}

The numerical results of the linear, SHG and PC optical conductivity for the $s$-wave tight-binding model are shown in Figs.~\ref{fig4}, \ref{fig5}, and \ref{fig6}, respectively. We set $t=1\times10^2$~meV, $q_x=0.02/a$, where $a\sim1$~\AA~is the lattice constant, $\mu=0.9\times10^2$~meV, $V_{\mathbf{k},\mathbf{k'}}=1\times10^2$~meV, $\Delta\approx 4.5$~meV is determined by a self-consistent gap equation, and $\eta = 10^{-1}$~meV. It is worth mentioning the choice of the magnitude of momentum $\mathbf{q}$. The current density is defined as $\mathbf{J}=e n_s \mathbf{v}_s$, where $n_s$ is the superfluid density that can be determined by $\text{Im}[\sigma(\omega \rightarrow 0)]=\lim_{\omega \rightarrow 0}n_s e^2/(m^* \omega)$. Therefore, $\mathbf{q}=m^* \mathbf{v}_s$ is estimated to be of the order of $10^{-2}/a$ if the current injected into the superconductor is about $1$~A and the two-dimensional (2D) superconducting film is $1$~mm$^2$ in area.

For the linear optical conductivity $\sigma^{(1)}(\omega)$ shown in Fig.~\ref{fig4}, its real part $\text{Re}[\sigma^{(1)}]$ is totally contributed from the bubble diagram in Fig.~\ref{fig3}(b) owing to the purely imaginary nature of the tadpole diagram in Fig.~\ref{fig3}(a). Similar to the results of previous research~\cite{papaj2022}, the real part shows a resonance peak near $\hbar\omega=2\Delta$, which is caused by supercurrent, and has $1/\sqrt{\hbar\omega-2\Delta}$ frequency dependence above the gap edge. Below the gap edge, there is no optical transition, and thus the real part vanishes. When vertex correction is included, the response is reduced in the full frequency range. The imaginary part $\text{Im}[\sigma^{(1)}]$ is mainly contributed from the tadpole diagram, which is three orders of magnitude larger than the bubble diagram. $\text{Im}[\sigma^{(1)}(\omega)]$ has $\omega^{-1}$ frequency dependence as shown in Fig.~\ref{fig4}(b), which reflects the Meissner effect. Meanwhile, $q$ has a negligible effect on $\text{Im}[\sigma^{(1)}(\omega)]$. Since the dominant contribution comes from the tadpole diagram, the vertex correction makes almost no difference and $\text{Im}[\sigma^{(1)}_{xx}(\omega)]$ is almost the same as $\text{Im}[\sigma^{(1)}_{yy}(\omega)]$.

\begin{figure}[t]
\begin{center}
\includegraphics[width=3.3in,clip=true]{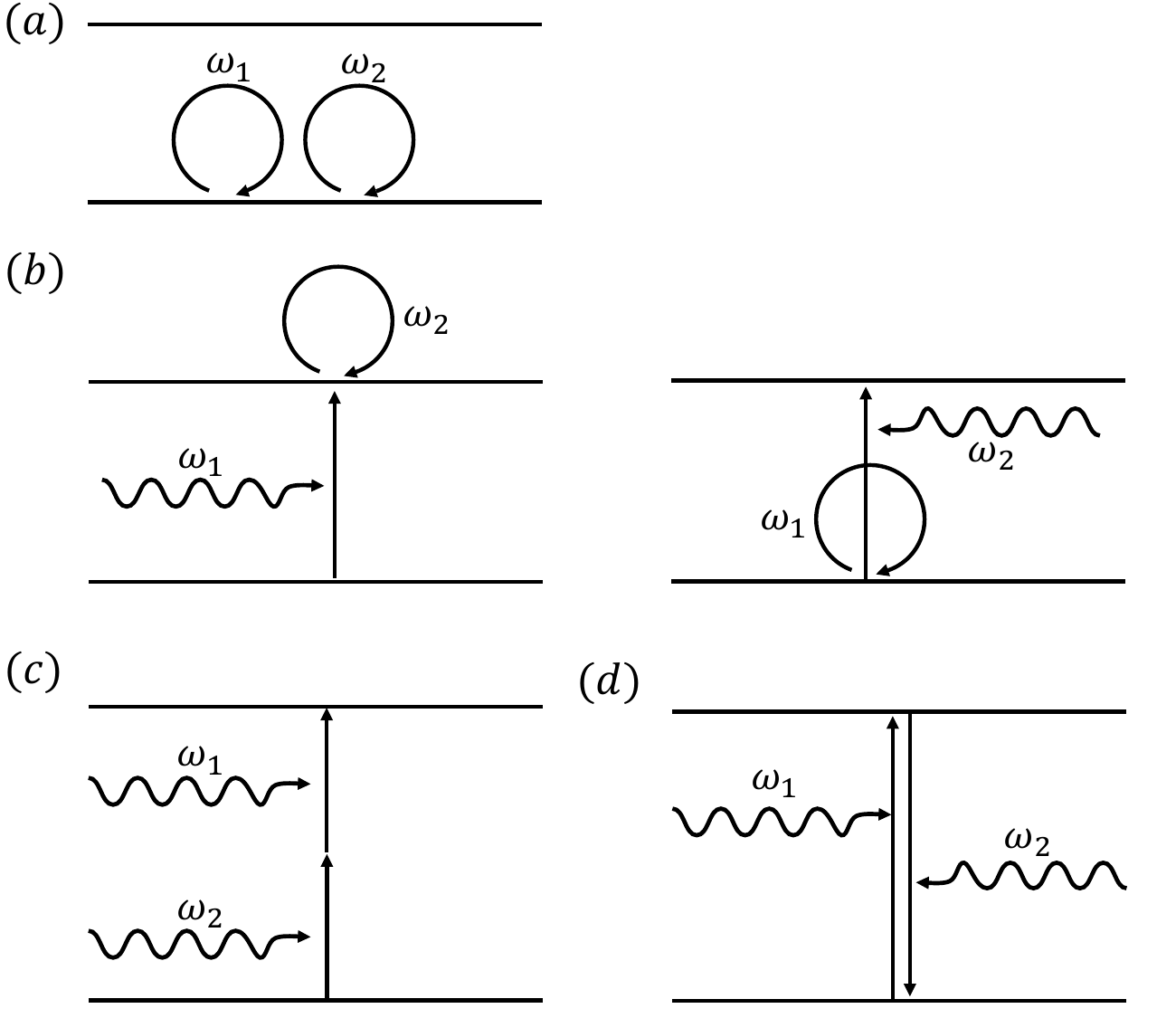}
\end{center}
\caption{Schematic picture for the second-order optical process. (a) Pure intraband response, and no resonant frequency appears. (b) Interband and intraband mixing process. In SHG and PC, resonance happens at $\hbar\omega=2\Delta$. (c),(d) Pure interband process. At $\hbar\omega=\Delta$, (c) resonance occurs in SHG, while (d) no resonance appears in PC.}
\label{fig7}
\end{figure}

For the second order in Figs.~\ref{fig5} and \ref{fig6}, $\text{Im}[\sigma^{(2)}]$ vanishes below the superconducting gap, since no quasiparticle optical transition occurs, while $\text{Re}[\sigma^{(2)}]\propto\omega^{-2}$ in the low-frequency limit. In the gapped $s$-wave superconductor with a dc supercurrent, both $\mathcal{T}$ and $\mathcal{I}$ symmetry are broken, but the combined $\mathcal{TI}$ symmetry is preserved, and the divergent behavior is attributed to the response of nonreciprocal superfluid density~\cite{watanabe2022a,watanabe2022b}. Namely, the divergence of $\text{Re}[\sigma^{(2)}]$ is robust against the choice of $\mathbf{q}\neq0$ and small parameter $\eta$ in the Green's function. In the SHG of Fig.~\ref{fig5}(b), there are two response peaks near $\hbar\omega=2\Delta$ and $\hbar\omega=\Delta$ in $\text{Im}[\sigma^{(2)}_{\text{SHG},xxx}]$, which arise from the bubble diagram in Figs.~\ref{fig3}(d) and \ref{fig3}(e). The former has a peak near $\hbar\omega=2\Delta$, and the latter has a peak near $\hbar\omega=\Delta$, which corresponds to and is very well explained by the optical transition processes in Figs.~\ref{fig7}(b) and \ref{fig7}(c), respectively. Remarkably, the current-induced peak in $\text{Im}[\sigma^{(2)}_{\text{SHG},xxx}]$ at $\hbar\omega=2\Delta$ is proportional to the square of the supercurrent density since it is mainly contributed from the bubble diagram of Fig.~\ref{fig3}(d). Since the bare vertex $\gamma_{xy}=0$, Fig.~\ref{fig3}(d) does not contribute to $\sigma^{(2)}_{\text{SHG},xyy}$, and thus there is no peak near $\hbar\omega=2\Delta$. Similarly, $\sigma^{(2)}_{\text{SHG},yyx}$ does not show the peak near $\hbar\omega=\Delta$ due to $\gamma_{yx}=0$ and no contribution from Fig.~\ref{fig3}(e). Meanwhile, in Fig.~\ref{fig6} for PC, only the resonant peak near $\hbar\omega=2\Delta$ may exist, as seen in $\sigma^{(2)}_{\text{PC},xxx}$ and $\sigma^{(2)}_{\text{PC},yyx}$. No peak will appear near $\hbar\omega=2\Delta$ in $\sigma^{(2)}_{\text{PC},xyy}$ for $\gamma_{xy}=0$. For both SHG and PC processes in the second-order response, the asymptotic behavior near $\hbar\omega=\Delta$ has a similar form to that near $\hbar\omega=2\Delta$. Namely, the frequency dependence of $\text{Re}[\sigma^{(2)}]$ shows $1/\sqrt{\Delta-\hbar\omega}$ and $1/\sqrt{2\Delta-\hbar\omega}$ when $\hbar\omega$ is below $\Delta$ and $2\Delta$, respectively, while $\text{Im}[\sigma^{(2)}]$ scales as $1/\sqrt{\hbar\omega-\Delta}$ and $1/{\sqrt{\hbar\omega-2\Delta}}$ above the resonance frequencies. After vertex correction, the entire magnitude of the second-order response is reduced, $\sigma^{(2)}_{xyy}$ and $\sigma^{(2)}_{yyx}$ even reverse sign, while the shape of the frequency-dependence curve is unchanged. Interestingly, we find that in this model, the contribution from the triangular diagram of Fig.~\ref{fig3}(f) can be neglected compared with tadpole (Fig.~\ref{fig3}(c)) and bubble (Fig.~\ref{fig3}(d) and~\ref{fig3}(e)) diagrams, where the latter reveals similar resonant behaviors and asymptotic frequency dependence to that of linear order. 

\subsection{Dirac fermion with $s$-wave pairing}
\label{dirac}

Then we proceed to study the Dirac fermion with an $s$-wave superconducting pairing, which describes the surface state of a 3D topological insulator with the proximity effect from a conventional $s$-wave superconductor~\cite{zhu2021}. In the presence of a dc supercurrent, the BdG Hamiltonian is
\begin{equation}
H^{\text{Dirac}}_{\text{BdG}}(\mathbf{k})= (\hbar v(k_x+q_x)s_y-\hbar vk_ys_x-\mu)\tau_z+\Delta\tau_x.
\end{equation}
Here, $v$ is the Fermi velocity, and we set $\mu$ staying high above the Dirac point. Thus we can focus on the upper Dirac bands around the proximity gap when the photon is in the terahertz frequency. The projected single-band model for the Dirac fermion is
\begin{equation}\label{diracfermion}
H^{\text{D}}_{\text{BdG}}(\mathbf{k})=
\hbar vq_x\hat{k}_x\tau_0+(\hbar vk-\mu)\tau_z+\Delta\tau_x.
\end{equation}
where $\hat{k}_x\equiv k_x/k$ and $k=\sqrt{k_x^2+k_y^2}$. The model has $C_{2x}$ symmetry,  similar to the above $s$-wave single-band model, and the symmetry-allowed nonzero independent components are $\sigma^{(1)}_{xx}, \sigma^{(1)}_{yy}$ for linear order, and $\sigma^{(2)}_{xxx},\sigma^{(2)}_{xyy},\sigma^{(2)}_{yyx}$ for second order.

\begin{figure}[b]
\begin{center}
\includegraphics[width=3.4in,clip=true]{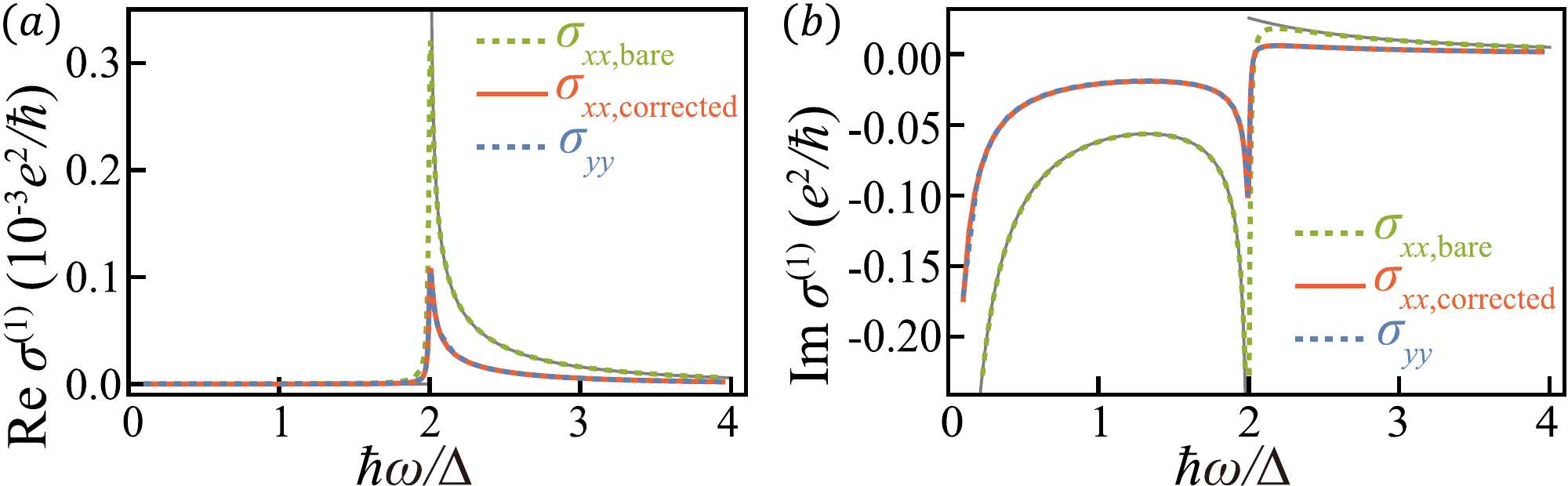}
\end{center}
\caption{$\sigma^{(1)}(\omega)$ of Dirac fermion model. Green (blue) dashed line denotes $\sigma^{(1)}_{xx}(\omega)$ without vertex correction ($\sigma^{(1)}_{yy}(\omega)$); $\sigma^{(1)}_{xx}(\omega)$ with vertex correction is shown in red.} (a) $\text{Re}[\sigma^{(1)}]$. (b) $\text{Im}[\sigma^{(1)}]$. Gray lines are the analytical result of $\sigma^{(1)}_{xx}(\omega)$ without vertex correction in Appendix~\ref{app C}.

\label{fig8}
\end{figure}

\begin{figure}[t]
    \begin{center}
    \includegraphics[width=3.4in,clip=true]{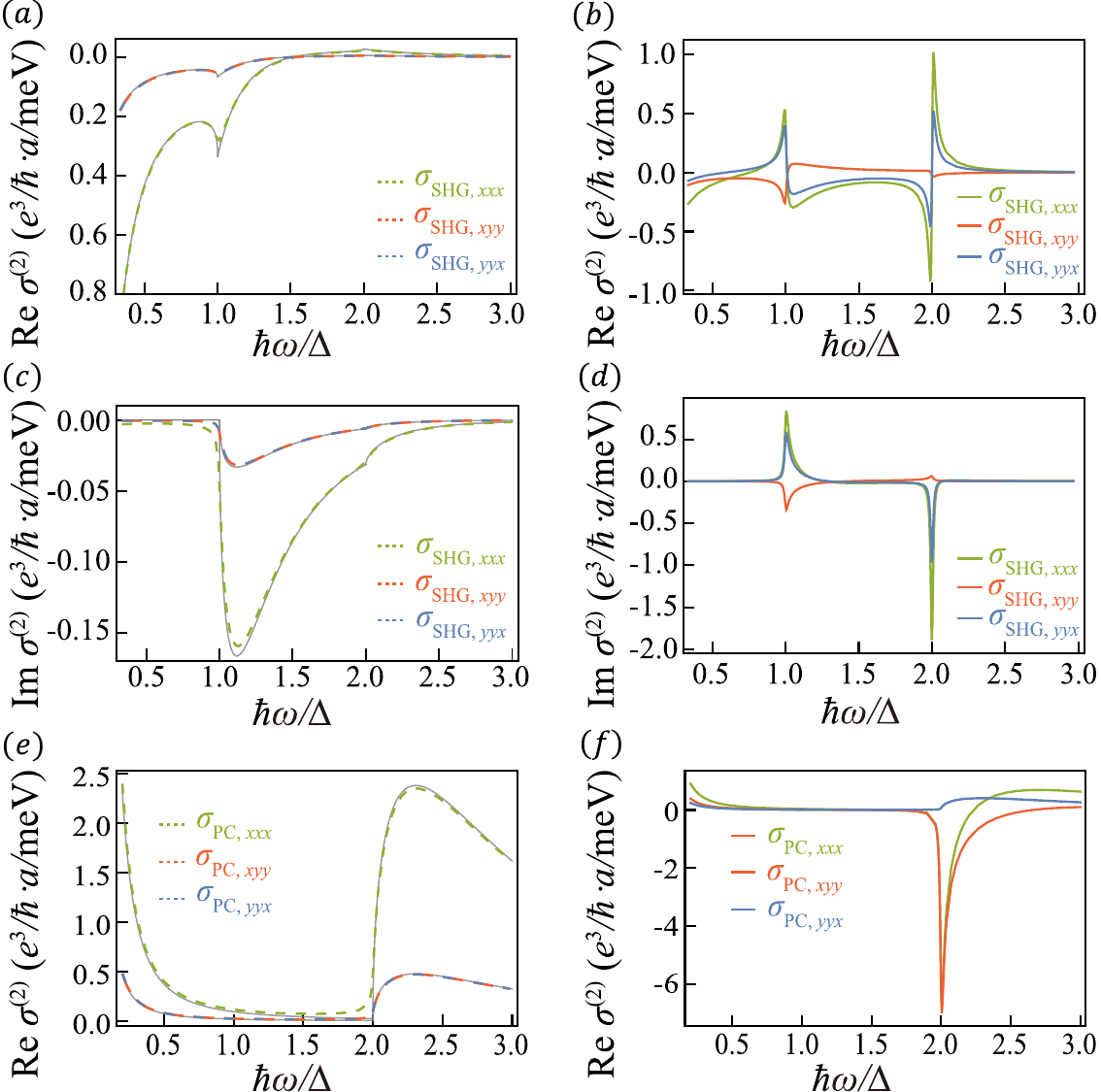}
    \end{center}
    \caption{$\sigma^{(2)}$ of Dirac fermion model. (a),(c) Real and imaginary part of $\sigma_{\text{SHG}}$ without vertex correction, respectively. Gray lines are the analytical result without vertex correction. (e) Real part of $\sigma_{\text{PC}}$ without vertex correction. The optical spectrum of $\sigma_{xyy}$ and $\sigma_{yyx}$ is the same. Corresponding results with vertex correction are shown in (b),(d) and (f), respectively. After vertex correction, $\sigma_{xyy}$ and $\sigma_{yyx}$ differ.}
    \label{fig9}
\end{figure}

The numerical results of linear, SHG, and PC optical conductivity for the Dirac fermion model are shown in Figs.~\ref{fig8} and \ref{fig9}, respectively. Meanwhile, for comparison, the analytic results for the responses without vertex correction are listed in Appendix~\ref{app C}. We set $q_x=0.005/a$, $\mu=2\times10^2$~meV, $V_{\mathbf{k},\mathbf{k'}}=0.8\times10^2$~meV, $\hbar v a= 6.6\times10^2$~meV, $\eta=5\times10^{-2}$~meV, and $\Delta\approx5$~meV. Due to the linear dispersion of the Dirac fermion, all vertex functions with more than one photon are zero. As a result, only the bubble diagram (Fig.~\ref{fig3}(b)) contributes to $\sigma^{(1)}$ and the triangular diagram (Fig.~\ref{fig3}(f)) contributes to $\sigma^{(2)}$. As shown in Figs.~\ref{fig8} and \ref{fig9}, in the low-frequency regime, $\text{Im}[\sigma^{(1)}]$ shows $\omega^{-1}$ behavior and $\text{Re}[\sigma^{(2)}]$ has $\omega^{-2}$ behavior. Below the superconducting gap, there is no optical transition, and $\text{Re}[\sigma^{(1)}]$ and $\text{Im}[\sigma^{(2)}]$ vanish. Without vertex correction, $\text{Re}[\sigma^{(1)}]$ has $1/\sqrt{\hbar\omega-2\Delta}$ asymptotic behaviors above the gap edge, and $\text{Im}[\sigma^{(1)}]$ scales as $1/\sqrt{2\Delta-\hbar\omega}$ below the gap edge [Fig.~\ref{fig8} green dashed line], which are the feature from the bubble diagram. However, for $\sigma^{(2)}$, no sharp peak appears, as shown in Figs.~\ref{fig9}(a),~\ref{fig9}(c) and~\ref{fig9}(e), which is the feature of the triangular diagram. When vertex correction is considered, the real and imaginary parts of $\sigma^{(1)}$ are reduced since both of them are purely contributed from the bubble diagram. However, the vertex correction does not change the resonance shape of the response curve in $\sigma^{(1)}$. In addition, the vertex corrected $\sigma_{xx}^{(1)}$ is identical to $\sigma_{yy}^{(1)}$, which is consistent with previous research~\cite{crowley2022}, while in the second-order response, besides the quantitative difference between the bare value and vertex corrected $\sigma^{(2)}$, the sharp peaks appear near $\hbar\omega=\Delta$ and $\hbar\omega=2\Delta$ with vertex correction.

\subsection{$d$-wave single-band superconductor}
\label{d-wave}

Finally, we study a different type of superconducting pairing, namely, a single-band spin-degenerate $d$-wave superconductor. The tight-binding model on a square lattice is
\begin{equation}\label{tightbinding for d}
H^{d}_{\text{TB}}(\mathbf{k})=
    \begin{pmatrix}
        \epsilon_{\mathbf{k}+q\hat{x}} & \Delta \phi_\mathbf{k} \\
        \Delta \phi_\mathbf{k} & -\epsilon_{-\mathbf{k}+q\hat{x}}
    \end{pmatrix},
\end{equation}
where $\epsilon_{\mathbf{k}}=t(2-\cos k_x-\cos k_y-\mu)$, $\phi_\mathbf{k}=(\cos{k_x}-\cos{k_y})$. The independent nonzero components of $\sigma^{(1)}$ and $\sigma^{(2)}$ are the same as in the $s$-wave model since the $d$-wave model here has $C_{2x}$ symmetry with the supercurrent flows along the $x$ axis. In the numerical calculations for $\sigma^{(1)}$ and $\sigma^{(2)}$, we set $q_x=0.07/a$,  $\mu=0.9\times10^2$~meV, $V=1.8\times10^2$~meV, $\eta=3\times10^{-2}$~meV, and $\Delta\approx23$~meV determined by Eq.~(\ref{gapeqn}). 

\begin{figure}[t]
\begin{center}
\includegraphics[width=3.4in,clip=true]{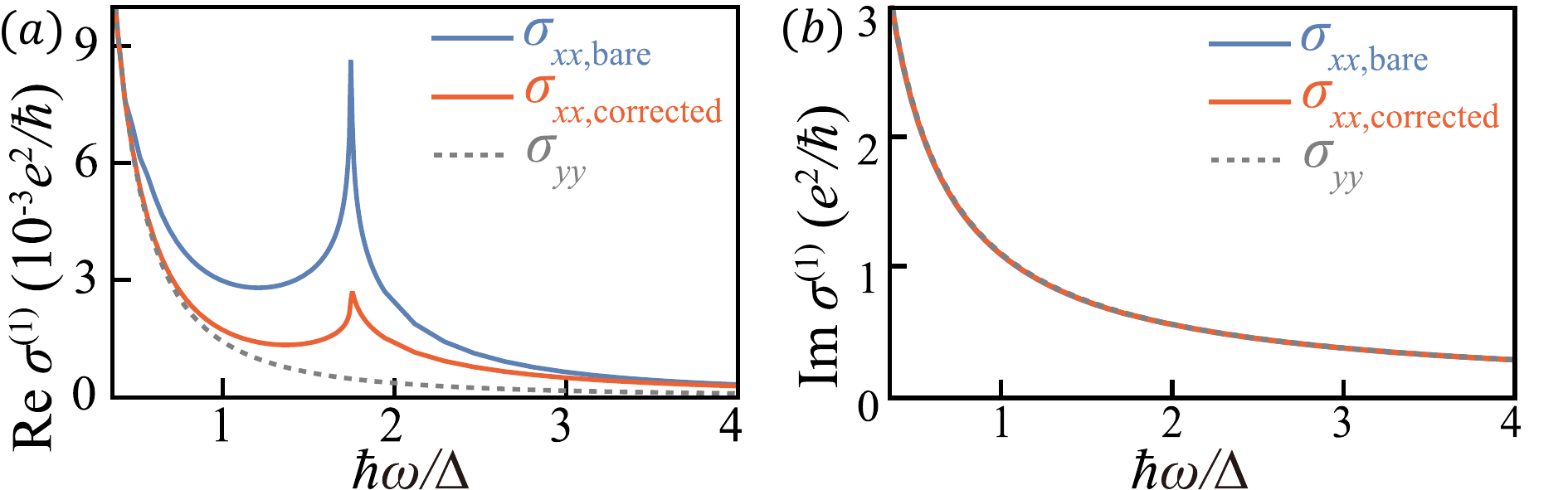}
\end{center}
\caption{$\sigma^{(1)}(\omega)$ of $d$-wave superconductor. (a) $\text{Re}[\sigma^{(1)}_{xx}(\omega)]$ without (blue solid line) and with (red solid line) vertex correction. (b) $\text{Im}[\sigma^{(1)}_{xx}(\omega)]$ without and with vertex correction. $\sigma^{(1)}_{yy}(\omega)$ (gray dashed line) is also shown.}
\label{fig10}
\end{figure}

\begin{figure}[b]
\begin{center}
\includegraphics[width=3.4in,clip=true]{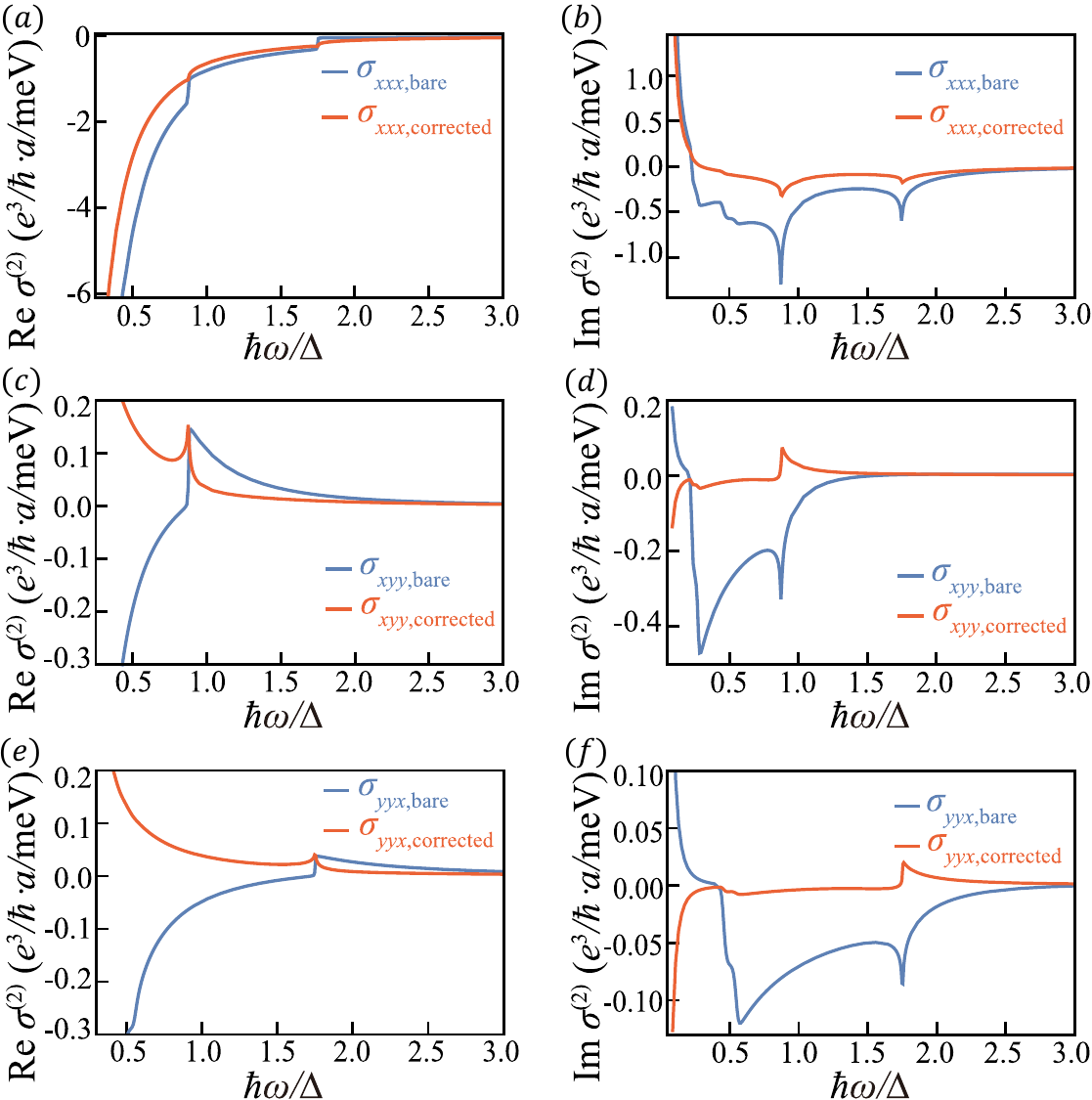}
\end{center}
\caption{$\sigma^{(2)}_{\text{SHG}}$ of $d$-wave superconductor. Blue and red lines show SHG without and with vertex correction, respectively. (a),(c),(e) Real part $\text{Re}[\sigma^{(2)}_{\text{SHG}}]$, (b),(d),(f) Imaginary part $\text{Im}[\sigma^{(2)}_{\text{SHG}}]$}
\label{fig11}
\end{figure}

Due to the nodal character of the quasiparticle spectrum, when a supercurrent is induced inside a $d$-wave superconductor, there exists a Bogoliubov Fermi surface which causes intraband conductivity. To capitalize the intraband response, we replace the factor $1/{\omega}$ in Eq.~(\ref{sigma1}) and $1/(\omega_1\omega_2)$ in Eq.~(\ref{sigma2}) by $1/({\omega+i\eta'})$ and $1/{(\omega_1+i\eta')(\omega_2+i\eta')}$, respectively~\cite{passos2018}. This corresponds to adiabatic switching the electrodynamic field as $\mathbf{A}(t)e^{-\eta' t}$, where we choose $\eta'=\eta$. Fig.~{\ref{fig10}} shows the linear optical conductivity with and without vertex correction. Near $\hbar\omega=2\Delta$, a peak induced by supercurrent appears in $\text{Re}[\sigma_{xx}^{(1)}]$, which is reduced when considering the vertex correction as expected. The intraband response gives a Drude peak in the low-frequency regime and scales as ${\eta'}/({\eta'^2+\omega^2})$. As with the $s$-wave superconductor, the dominant contribution to $\text{Im}[\sigma^{(1)}]$ comes from the tadpole diagram that accounts for the Meissner effect. 

\begin{figure}[t]
\begin{center}
\includegraphics[width=3.4in,clip=true]{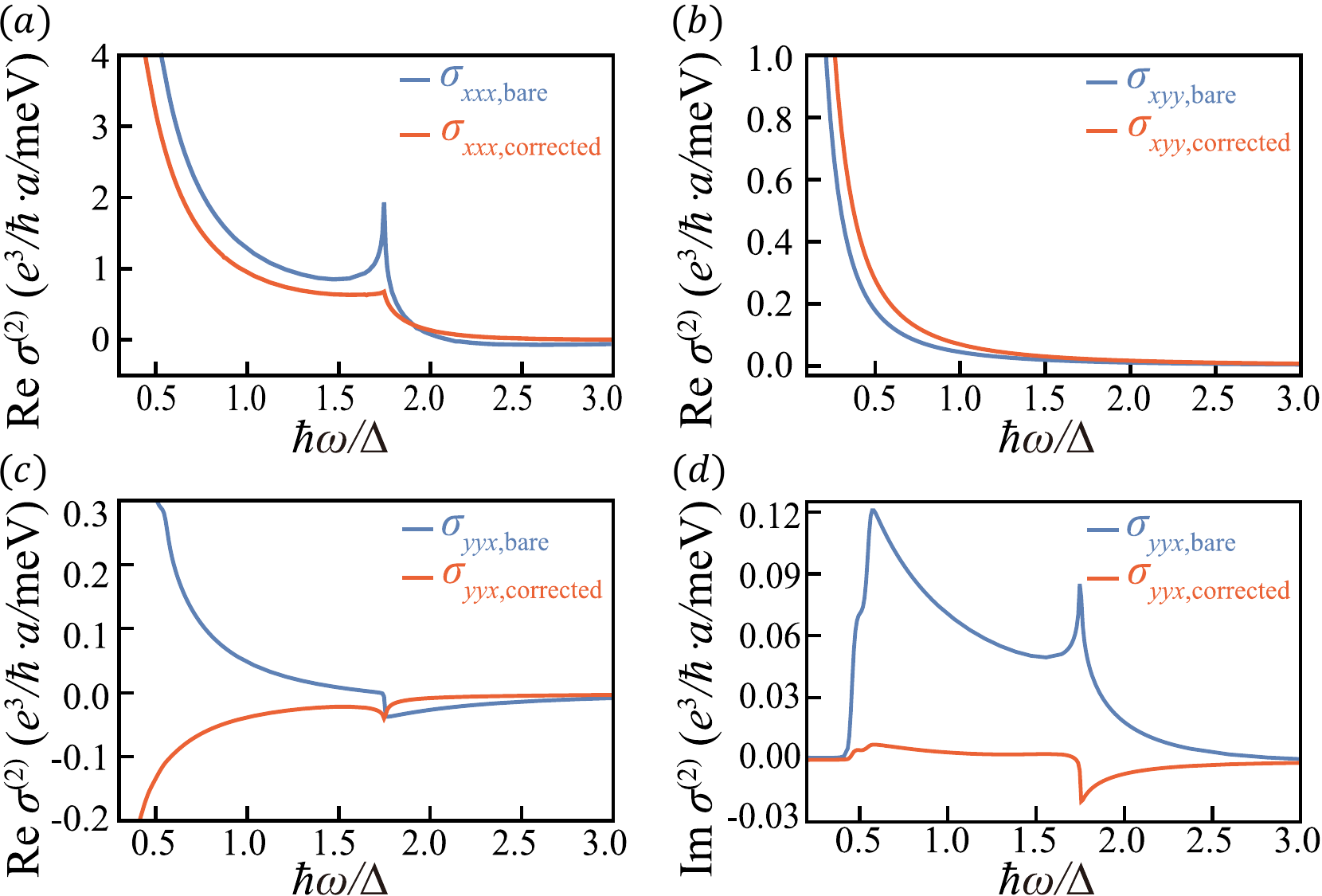}
\end{center}
\caption{$\sigma^{(2)}_{\text{PC}}$ of $d$-wave tight-binding model. Blue and red lines show PC without and with vertex correction, respectively. (a) $\text{Re}[\sigma^{(2)}_{\text{PC},xxx}]$, (b) $\text{Re}[\sigma^{(2)}_{\text{PC},xyy}]$, where both of their imaginary parts are zero. (c) $\text{Re}[\sigma^{(2)}_{\text{PC},yyx}]$. (d) $\text{Im}[\sigma^{(2)}_{\text{PC},yyx}]$.}
\label{fig12}
\end{figure}

Figs.~{\ref{fig11}} and {\ref{fig12}} show second-order processes SHG and PC, respectively. The real parts of $\sigma_{\text{SHG}}$ and $\sigma_{\text{PC}}$ show $\omega^{-2}$ divergent behavior, while the imaginary part $\text{Im}[\sigma^{(2)}]$ has a finite contribution from the intraband response in the low-frequency region, which is different from the $s$-wave case where $\text{Im}(\sigma^{(2)})$ vanishes due to the superconducting gap. Therefore, the current-induced second-order optical effect could distinguish different superconducting order parameters. Also different from the $s$-wave case, there are some kinks in $\text{Im}[\sigma^{(2)}]$ of the $d$-wave superconductor when $\hbar\omega/\Delta\sim0.5$ (i.e. $\hbar\omega\sim10$~meV), which corresponds to the optical transition happening near the Bogoliubov Fermi surface, and may come from Van Hove singularities. In general, the vertex correction will reduce the magnitude of the second-order response and some components even reverse sign, while with one exception $\sigma^{(2)}_{\text{PC},xyy}$ gets enhanced.

\section{Discussion}
\label{discussion}

We compare the magnitude of the predicted effect to the present experiments on second-order optical conductivity of superconductors. A current-enabled SHG was recently reported in the experimental of Ref.~\cite{nakamura2020}. In that experiment, the observed SHG nonlinear susceptibility attributed to the supercurrent flow in the NbN superconductor thin film can reach $\chi^{(2)}_{\text{ex}}\sim2.2\times10^8$~pm/V near the resonance frequency $\hbar\omega=2\Delta$. Our calculation of the 2D $s$-wave tight-binding model shows a maximum of $\text{Im}[\sigma^{(2)}_{\text{SHG},xxx}]$ and is about $7.2\times10^{-13}$~$\text{m}\cdot(\mathrm{\Omega}\cdot\text{V})^{-1}$; the 3D SHG conductivity can be obtained by the 2D value multiplied by the measured Fermi wave vector of NbN, $k_F\approx1.45$~$\text{\AA}^{-1}$~\cite{papaj2022,chockalingam2008}. Converted to susceptibility, the final result is $\chi^{(2)}_{\text{th}}\approx3.9\times10^8$~pm/V, which matches well with the experimental result. Meanwhile, the current-induced peak in $\text{Im}[\sigma^{(2)}(\omega)]$ is predicted to be proportional to the square of the supercurrent density, which is also consistent with the experimental observations~\cite{nakamura2020}. All of these mean that the proposed effect may have very well already been observed. However, more detailed studies are required to convincingly separate our mechanism from one based on vortex dynamics in that experiment.

So far, we only consider a clean system, and the effect of impurity scattering can be considered as a positive imaginary parameter into the Green's function. Thus the resulting response from both linear and second order is reduced, especially near the resonance frequency region, and the divergent behavior in the low-frequency limit is replaced by approaching to a large value. In the dirty limit, the linear response will return to Mattis-Bardeen theory~\cite{mattis1958}, which is also similar in the second-order effect; the optical response near the gap edge will be reduced since a small supercurrent does not dramatically change the band structure. Recently, the second-order optical effect of the diffusive superconductor has been studied~\cite{derendorf2023}.

In summary, we have studied a supercurrent-enabled second-order optical response in BCS superconductors of the clean limit. Such a supercurrent flow can be induced and controlled by an external magnetic field through the Meissner effect. Similar to the linear order case, we find a large second-order response near the gap edge in the presence of dc supercurrent. After vertex correction, the entire magnitude of the linear- and second-order responses is reduced, only some second-order components even reverse sign, while the shape of the frequency-dependence curve is unchanged. Here we point out that the contribution from collective modes has already been included in vertex correction reflecting the BCS electron-electron interactions. For the above several examples with typical superconducting order parameters, we show that the current-enabled second-order optical conductivity strongly depends on the type of superconducting pairing as well as the nature of the normal state. As such, the supercurrent-induced nonlinear optical spectroscopy provides a valuable toolbox to explore novel superconductors. Moreover, although we only present the nonlinear response up to second order in this manuscript, the diagrammatic method with vertex correction can be readily extended to a higher-order optical response. 

{\it Note added}: During the finalization of our manuscript, we learned of an independent work on the second-order optical Hall response of superconductors~\cite{parafilo2023}. They consider the case of weak disorder, which is different from our case.

\begin{acknowledgments}
This work is supported by the National Key Research Program of China under Grant No.~2019YFA0308404, the Natural Science Foundation of China through Grant No.~12174066, Science and Technology Commission of Shanghai Municipality under Grants No. 20JC1415900 and No. 23JC1400600, the Innovation Program for Quantum Science and Technology through Grant No.~2021ZD0302600, and Shanghai Municipal Science and Technology Major Project under Grant No.~2019SHZDZX01.
\end{acknowledgments}

\appendix

\section{Ward-Takahashi identity}\label{app A}
\begin{figure}[t]
\begin{center}
\includegraphics[width=3.0in,clip=true]{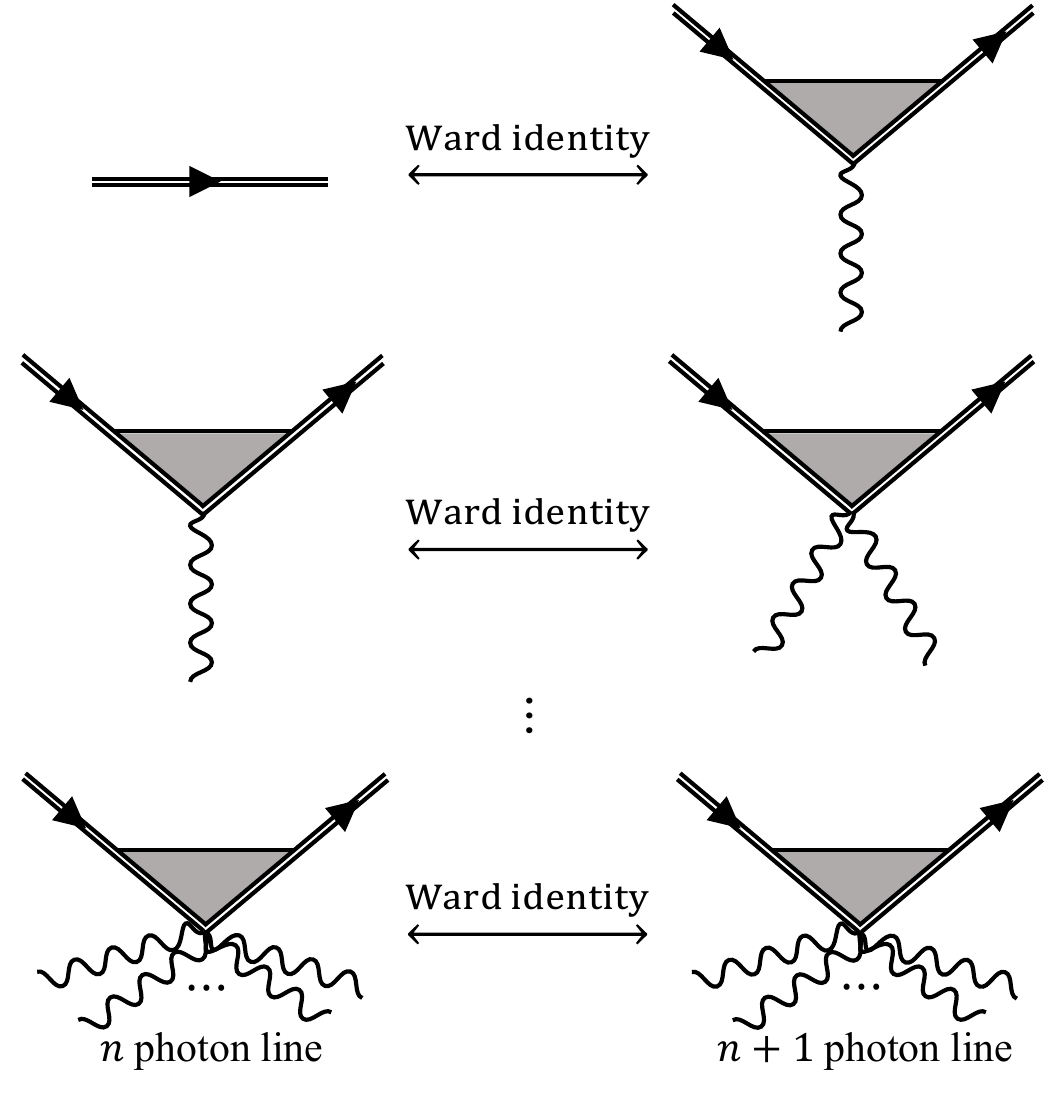}
\end{center}
\caption{Diagrammatical description of Ward identities. The $n$-photon vertex correction term has a quantitative relation to the $n+1$-photon vertex correction term.}
\label{fig13}
\end{figure}

Here we derive the Ward-Takahashi identity in a superconducting system, and apply it into the optical response to maintain the gauge invariance, which leads to vertex correction. The partition function is
\begin{equation}
    Z=\int \mathcal{D}\bar{\Psi}\mathcal{D}\Psi e^{-S[\bar{\Psi},\Psi,A]} .
\end{equation}
where $S$ is the action for the superconductor without electromagnetic field, and $ \Psi=(\psi_{\uparrow},\bar{\psi}_\downarrow)^T $ is the Nambu bi-spinor. The generating functional of the connected Green's function is
\begin{equation}
    G=\ln \int \mathcal{D}\bar{\Psi}\mathcal{D}\Psi e^{-S[\bar{\Psi},\Psi,A]-\int \mathrm{d}\mathbf{r}(J_{A_\mu}A_\mu +J_{\bar{\Psi}} \Psi + J_{\Psi} \bar{\Psi})},
\end{equation}
where $ J_{A_\mu}, J_{\bar{\Psi}}$ and $ J_\Psi$ are auxiliary fields. The expectation value of $ A_\mu, \bar{\Psi} $, and $ \Psi $ can be calculated through the generating functional,
\begin{equation}
  \langle A_\mu \rangle = \frac{\delta G}{\delta J_{A_\mu}},\langle \bar{\Psi} \rangle = \frac{\delta G}{\delta J_\Psi},\langle \Psi \rangle = \frac{\delta G}{\delta J_{\bar{\Psi}}}.
\end{equation}
The generating functional of the irreducible vertex function is obtained via Legendre transformation: $ \Gamma=G-\int \mathrm{d}\mathbf{r}\left(\langle A_\mu \rangle J_{A_\mu}+\langle \Psi \rangle J_{\bar{\Psi}}+\langle \bar{\Psi} \rangle J_\Psi\right) $. Taking the functional derivative of $\Gamma$ will give vertex functions,
\begin{equation}
    J_{A_\mu} = \frac{\delta \Gamma}{\delta \langle A_\mu \rangle},J_{\bar{\Psi}} = \frac{\delta \Gamma}{\delta \langle \Psi \rangle}, J_\Psi = \frac{\delta \Gamma}{\delta \langle \bar{\Psi} \rangle}.
\end{equation}
Gauge invariance requires that $ G $ is invariant under infinitesimal $ U(1) $  transformation: $ \Psi'=(\tau_0+i\theta \tau_3)\Psi, \bar{\Psi}'=(\tau_0-i\theta \tau_3)\bar{\Psi}, A'_\mu=A_\mu-\partial_\mu \theta$. From $\delta G/{\delta \theta}=0 $, we have
\begin{equation} \label{ward0}
    \partial_{x^\mu}\left(\frac{\delta \Gamma}{\delta \langle A_\mu \rangle}\right)=i\left(\frac{\delta \Gamma}{\delta \langle \Psi \rangle}\tau_3 \Psi - \bar{\Psi} \frac{\delta \Gamma}{\delta \langle \bar{\Psi} \rangle}\right).
\end{equation}
The series of Ward identities can be derived from Eq.~(\ref{ward0}) by taking the functional derivative of both sides with respect to the expectation value of the fields. For example, taking the functional derivative with respect to $ \langle \bar{\Psi} \rangle $ and $\langle \Psi \rangle$ gives
\begin{eqnarray}\label{ward1}
    &\partial_{x^\mu}& \frac{\delta^3 \Gamma}{\delta\langle \bar{\Psi}(z)\rangle \delta \langle \Psi(y) \rangle \delta \langle A_\mu (x) \rangle} = i \left[\frac{\delta^2 \Gamma}{\delta\langle \bar{\Psi}(z)\rangle \delta\langle\Psi(x)\rangle}\right.
    \nonumber
    \\
    &&\left.\times\tau_3 \delta (x-y)-\tau_3 \frac{\delta^2 \Gamma}{\delta \langle \bar{\Psi}(x) \rangle \delta \langle \Psi(y) \rangle} \delta(x-z)\right].
\end{eqnarray}
Converted to momentum representation, Eq.~(\ref{ward1}) becomes the Ward-Takahashi identity $ q_\mu \Gamma^1_\mu(k+q,k)=\tau_3 G^{-1}(k)-G^{-1}(k+q)\tau_3$, which gives the connection between the Green's function and the one-photon vertex function. The relation between the self-energy correction term $\Sigma(k)$ and one-photon vertex correction term $\Lambda^1_\mu(k+q,k)$ can be directly obtained by the Ward-Takahashi identity
\begin{equation}
    q_\mu \Lambda^1_\mu(k+q,k)=\Sigma(k+q) \tau_3 - \tau_3 \Sigma(k).
\end{equation}
In addition, if we further take the functional derivative of Eq.~(\ref{ward1}) with respect to $\langle A_\mu\rangle $, we will get the relation between the one-photon vertex correction term $\Lambda^1_\mu(k+q,k)$ and the two-photon vertex correction term $\Lambda^2_{\mu,\nu}(k+q'+q,k)$,
\begin{equation}
    q_\mu \Lambda^2_{\mu,\nu}(k+q'+q,k)=\Lambda^1_\mu(k+q'+q,k) \tau_3 - \tau_3 \Lambda^1_\mu(k+q',k)
\end{equation}
A similar procedure can be done to get the relation between the $n$-photon vertex correction term and the $n+1$-photon vertex correction term,
\begin{equation}
    q_\mu \Lambda^{n+1}_{\mu,\nu}(k+q'+q,k)=\Lambda^n_\mu(k+q'+q,k) \tau_3 - \tau_3 \Lambda^n_\mu(k+q',k)
\end{equation}
A diagrammatical description is shown in Fig.~\ref{fig13}, which needs to be satisfied when calculating the higher-order optical response of a superconductor. In this sense, the self-energy correction term $\Sigma(k)$ can be viewed as the zero-photon vertex correction term $\Lambda^0(k)$. 

\section{Vertex correction for optical response}\label{app B}
Here we give the detailed calculation of the vertex correction in the single-band superconductor models, then apply it to the optical response.

The $\gamma_{a_1,a_2,\cdots,a_n}$ matrix can be decomposed into two components in terms of Pauli matrices: $\gamma_{a_1,a_2,\cdots,a_n}=\gamma_{a_1,a_2,\cdots,a_n}^{(1)}\tau_0 + \gamma_{a_1,a_2,\cdots,a_n}^{(2)}\tau_3$. Similarly, the corrected vertex, $\Gamma$ matrix can be expanded as,
\begin{equation}\label{B1}
    \Gamma_{a_1,a_2,\cdots,a_n}=\gamma_{a_1,a_2,\cdots,a_n}+\sum_{i=0}^{3}\Gamma_{a_1,a_2,\cdots,a_n}^{(i)} \tau_i.
\end{equation} 
In the $s$-wave case, $\Gamma(p+q,p)$ is independent of $p$, and the photon momentum $\mathbf{q}$ is negligible, thus all expansion coefficients of the $\Gamma$ matrix can be solved by self-consistent Eq.~(\ref{Gamma}) at every frequency of incident light. 

To find out the resonance frequency clearly, we sum over the imaginary temporal component and reexpress the formulas of $\sigma^{(1)}$ and $\sigma^{(2)}$. The diagram in Fig.~\ref{fig3}(a) gives
\begin{equation}\label{B2}
    \int[\mathrm{d}\mathbf{k}]\left(-\gamma_{ab}^{(1)} + \frac{\bar{\epsilon}_{\mathbf{k}} \gamma_{ab}^{(2)}}{E'_{\mathbf{k}}}\right),
\end{equation}
where we denote $\bar{\epsilon}_{\mathbf{k}} =(\epsilon_{\mathbf{k}+\mathbf{q}}+\epsilon_{-\mathbf{k}+\mathbf{q}})/2$, $E'_{\mathbf{k}}=\sqrt{\bar{\epsilon}_{\mathbf{k}}^2+\Delta^2}$ and abbreviate the integral over $\mathbf{k}$ to $\int[\mathrm{d}\mathbf{k}]$. Fig.~\ref{fig3}(b) gives
\begin{eqnarray}\label{B3}
    \int[\mathrm{d}\mathbf{k}]&&\frac{2\Delta}{E'_{\mathbf{k}}[4 {E'_{\mathbf{k}}}^2-(\hbar\tilde\omega)^2]}  \left[-2\bar{\epsilon}_{\mathbf{k}}\Gamma_b^{(1)}(\omega)\gamma_a^{(2)}\right.
    \\
    &&\left.+i\hbar\tilde\omega\Gamma_b^{(2)}(\omega)\gamma_a^{(2)}+2\Delta(\gamma_a^{(2)}\gamma_b^{(2)}+\Gamma_b^{(3)}(\omega)\gamma_a^{(2)})\right]\nonumber
\end{eqnarray}
where $\tilde\omega=\omega+i0^+$. The denominator indicates that the resonance frequency is $\hbar\omega\approx 2\Delta$.
Eq.~(\ref{sigma2}) consists of four parts, each is represented by a diagram in Figs.~\ref{fig3}(e) and \ref{fig3}(f). The first part reads
\begin{equation}\label{B4}
    \int[\mathrm{d}\mathbf{k}]\left(\gamma_{abc}^{(1)} + \frac{\bar{\epsilon}_{\mathbf{k}} \gamma_{abc}^{(2)}}{E'_{\mathbf{k}}}\right).
\end{equation}
The second part is a bubble diagram,
\begin{eqnarray}\label{B5}
    \int[\mathrm{d}\mathbf{k}] &&\frac{-2\Delta}{E'_{\mathbf{k}}[4{E'_{\mathbf{k}}}^2-(\hbar\tilde\omega_2)^2]} \left[-2\bar{\epsilon}_{\mathbf{k}}\Gamma_c^{(1)}(\omega_2)\gamma_{ab}^{(2)}\right.
    \nonumber
    \\
    &&+i\hbar\tilde\omega_2 \Gamma_c^{(2)}(\omega_2)\gamma_{ab}^{(2)}\left.+2\Delta(\gamma_{ab}^{(2)}\gamma_c^{(2)}+\Gamma_c^{(3)}(\omega_2)\gamma_{ab}^{(2)})\right]
    \nonumber
    \\
    &&+(b,\omega_1)\leftrightarrow(c,\omega_2)
\end{eqnarray}
The denominator indicates that the resonance frequency is $\hbar\omega\approx 2\Delta$. The third part is another bubble diagram,
\begin{eqnarray}\label{B6}
    &&\int[\mathrm{d}\mathbf{k}] \frac{-2\Delta}{E'_{\mathbf{k}}[4{E'_{\mathbf{k}}}^2-(\hbar\tilde\omega_{12})^2]} \left[-2\bar{\epsilon}_{\mathbf{k}}\Gamma_c^{(1)}(-\omega_{12})\gamma_{bc}^{(2)}\right.
    \\
    &&\left.-i\hbar\tilde\omega_{12}\Gamma_c^{(2)}(-\omega_{12})\gamma_{bc}^{(2)}+2\Delta(\gamma_{bc}^{(2)}\gamma_a^{(2)}+\Gamma_c^{(3)}(-\omega_{12})\gamma_{bc}^{(2)})\right]\nonumber
\end{eqnarray}
where $\omega_{12}=\omega_1+\omega_2$ and $\tilde\omega_{12}=\tilde\omega_1+\tilde\omega_2$. The denominator indicates that in the case of SHG, the resonance frequency is $\hbar\omega\approx\Delta$, and in the case of photocurrent effect, no resonance frequency appears.

The triangular diagram is far more complicated; we only show the result without vertex correction,
\begin{equation}\label{B7}
    \begin{aligned}
    &\int[\mathrm{d}\mathbf{k}]\frac{16 \Delta^2 \bar{\epsilon}_{\mathbf{k}}[12 {E'_{\mathbf{k}}}^2-(\hbar\tilde\omega_{12})^2+\hbar\tilde\omega_1\hbar\tilde\omega_2]\gamma_{a}^{(2)} \gamma_{b}^{(2)} \gamma_{c}^{(2)}}{E'_{\mathbf{k}}[4{E'_{\mathbf{k}}}^2-(\hbar\tilde\omega_{12})^2]
    [4{E'_{\mathbf{k}}}^2-(\hbar\tilde\omega_1)^2][4{E'_{\mathbf{k}}}^2-(\hbar\tilde\omega_2)^2]}
    \end{aligned}
\end{equation}
Likewise, the denominator indicates that for SHG, the resonance frequencies are $\hbar\omega\approx\Delta$ and $\hbar\omega\approx2\Delta$; for the photocurrent effect, the resonance frequency is $\hbar\omega\approx 2\Delta$. Vertex correction will modify this term significantly, while leaving the resonance frequencies unchanged, which can be seen in the Dirac fermion case in the main text.

As a side note, if there is no supercurrent, $q=0$, $\gamma^{(2)}_{a}=\gamma^{(1)}_{ab}=\gamma^{(2)}_{abc}=0$, then Eq.~(\ref{B3})-(\ref{B6}) and the real part of Eq.~(\ref{B2}) vanish in the condition of $\epsilon_\mathbf{k}=\epsilon_{-\mathbf{k}}$. Only the imaginary part of $\sigma^{(1)}$ survives and there is no optical transition at any finite frequency, verifying the effect of the supercurrent. Besides, the zero response of the second order reflects inversion symmetry of the system when $q=0$.

\begin{figure}[t]
    \begin{center}
    \includegraphics[width=3.3in,clip=true]{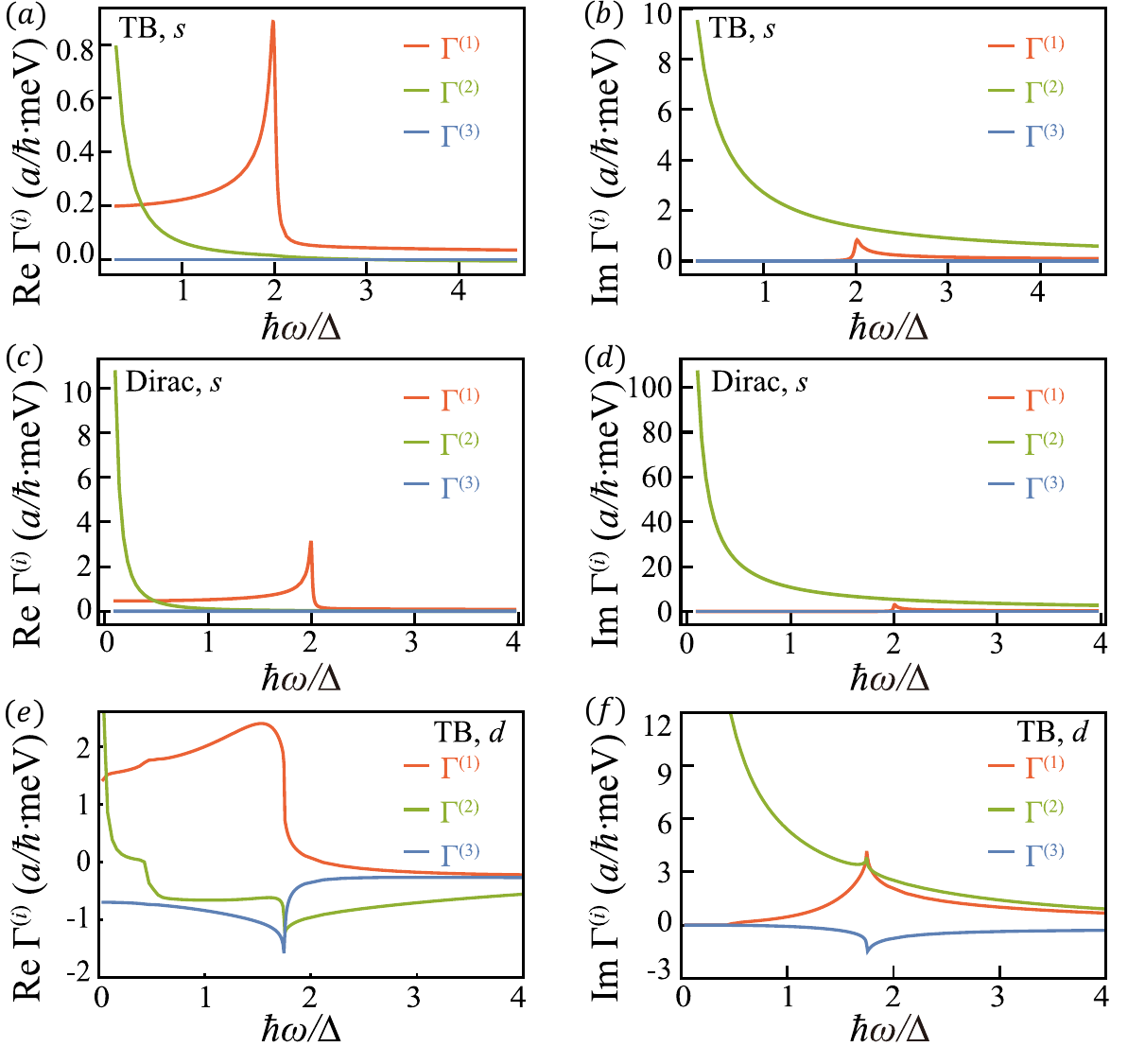}
    \end{center}
    \caption{Components of $\Gamma$ vs frequency. (a),(b) Calculated from $s$-wave tight-binding model; (c),(d) Calculated from Dirac fermion model; (e),(f) Calculated from $d$-wave tight-binding model.}
    \label{fig14}
\end{figure}

In the $d$-wave case, $ V_{\mathbf{k},\mathbf{k'}} $ can factorized as: $ V \phi_\mathbf{k}\phi_\mathbf{k'} $, where $V$ is a constant and $\phi_\mathbf{k}=(\cos{k_x}-\cos{k_y})$, and the self-consistent equation for the corrected vertex function becomes
\begin{equation}\label{Gamma for d}
    \begin{aligned}
        & \frac{\Gamma(p+q,p)}{\phi_\mathbf{p}} = \frac{\gamma(p+q,p)}{\phi_\mathbf{p}}+ V\frac{1}{\beta}\sum_{k_0} \int \frac{\mathrm{d^2}\mathbf{k}}{(2\pi)^2}\\
        & [\phi_\mathbf{k} \tau_3 G(k_0+q_0,\mathbf{k}+\mathbf{q}) \Gamma(k+q,k) G(k_0,\mathbf{k})\tau_3] 
    \end{aligned}
\end{equation}
After being divided by $\phi_\mathbf{p}$, the integrand above is independent of $\mathbf{p}$, and thus we can expand the $\Gamma$ matrix as:
\begin{equation}
    \Gamma(p+q,p) = \gamma(p+q,p) + \phi_\mathbf{p}  \sum_{i=0}^{3} \Gamma^{(i)} \tau_i
\end{equation}
In this manner, all the expansion coefficients of the $\Gamma$ matrix can be solved through Eq.~(\ref{Gamma for d}) for every frequency (again, we consider $\mathbf{q}=0$); then the optical responses of the $d$-wave superconductor can be obtained by Eqs.~(\ref{sigma1}) and (\ref{sigma2}). 

We show $\Gamma^{(1)},\Gamma^{(2)},\Gamma^{(3)}$ with respect to frequency calculated from the $s$-wave tight-binding model, Dirac fermion model and $d$-wave tight-binding model in Fig.~\ref{fig14}. Clearly, in the first two models, $\Gamma^{(1)}$ has a peak near resonance frequency, while $\Gamma^{(2)}$ is almost proportional to $\omega^{-1}$, and $\Gamma^{(3)}$ can be neglected compared with $\Gamma^{(1)}$ and $\Gamma^{(2)}$. We found that the $\Gamma^{(1)}$ components contribute little to modification of the bubble diagram, but are important to modification of the triangular diagram, and thus can change the second order conductivity distinctly in the Dirac fermion model. In the $d$-wave case, besides the peaks near $\hbar\omega=2\Delta$ and $\omega^{-1}$ divergence of $\Gamma^{(2)}$ at the low-frequency region, more complex behaviors appear, especially near the frequency region when an optical transition occurs near the Bogoliubov Fermi surface that is induced by the supercurrent. These behaviors are reflected in the intricate response after vertex correction of the $d$-wave superconductor. 

\section{Analytical result for optical responses of Dirac fermion with $s$-wave pairing}\label{app C}
\begin{widetext}
The optical conductivity for the Dirac fermion with $s$-wave superconducting pairing without vertex correction can be calculated analytically. The real part of $\sigma^{(1)}_{xx}$ is
\begin{equation}
\text{Re}(\sigma^{(1)}(\omega))=\frac{e^2}{\hbar} 3\pi^2 \frac{\Delta}{\mu} \frac{(\hbar v q)^2}{(\hbar \omega)^2} \frac{\Delta}{\sqrt{(\hbar \omega)^2-4\Delta^2}} \Theta[\hbar \omega-2\Delta]
\end{equation}
and the imaginary part can be obtain via Kramers-Kronig (K-K) relation as
\begin{equation}
    \text{Im}(\sigma^{(1)}(\omega))=\frac{e^2}{\hbar}6\pi\frac{\Delta}{\mu} \frac{(\hbar v q)^2}{(\hbar \omega)^2}
    \begin{cases}
    -\frac{\Delta\arcsin{(\frac{\hbar\omega}{2\Delta})}}{\sqrt{4\Delta^2-(\hbar \omega)^2}} & 0<\hbar \omega<2\Delta \\
    \frac{\Delta\operatorname{arcosh}{(\frac{\hbar\omega}{2\Delta})}}{\sqrt{(\hbar \omega)^2-4\Delta^2}} & \hbar\omega>2\Delta
    \end{cases}
\end{equation}
where $\sigma^{(1)}_{yy}(\omega)=\sigma^{(1)}_{xx}(\omega)/3$ due to the difference between $\gamma_x$ and $\gamma_y$ in angular integrations.

The imaginary part of $\sigma^{(2)}_{\text{SHG},xxx}(\omega)$ is
\begin{equation}
    \begin{aligned}
    \text{Im}[\sigma^{(2)}_{\text{SHG},xxx}(\omega)]=-\frac{e^3}{\hbar^2} 5\pi^2 \frac{\Delta^2}{\mu^2} \frac{(\hbar v q)^3}{(\hbar\omega)^5} \frac{\hbar v}{\mu}
    \left(\sqrt{(\hbar\omega)^2-\Delta^2}\Theta[\hbar\omega-\Delta]-\sqrt{(\hbar\omega)^2-4\Delta^2}\Theta[\hbar\omega-2\Delta]\right)
    \end{aligned}
\end{equation}
Through the generalized K-K relation for the nonlinear response~\cite{bassani1992}, the real part is given by
\begin{equation}
    \begin{aligned}
    \text{Re}[\sigma^{(2)}_{\text{SHG},xxx}(\omega)]=\frac{e^3}{\hbar} 10\pi \frac{\Delta^2}{\mu^2} \frac{(\hbar v q)^3}{(\hbar\omega)^5} \frac{\hbar v}{\mu}
    \begin{cases}
    \sqrt{\Delta^2-(\hbar\omega)^2}\arcsin{(\frac{\hbar\omega}{\Delta})}-\sqrt{4\Delta^2-(\hbar\omega)^2}\arcsin{(\frac{\hbar\omega}{2\Delta})} & 0<\hbar\omega<\Delta \\
    \sqrt{(\hbar\omega)^2-\Delta^2}\operatorname{arcosh}{(\frac{\hbar\omega}{\Delta})}-\sqrt{4\Delta^2-(\hbar\omega)^2}\arcsin{(\frac{\hbar\omega}{2\Delta})}&\Delta<\hbar\omega<2\Delta\\
    \sqrt{(\hbar\omega)^2-\Delta^2}\operatorname{arcosh}{(\frac{\hbar\omega}{\Delta})}-\sqrt{(\hbar\omega)^2-4\Delta^2}\operatorname{arcosh}{(\frac{\hbar\omega}{2\Delta})}&\hbar\omega>2\Delta
    \end{cases}
    \end{aligned}
\end{equation}
$\sigma^{(2)}_{\text{PC},xxx}(\omega)$ is purely real:
\begin{equation}
\sigma^{(2)}_{\text{PC},xxx}(\omega)=\frac{e^3}{\hbar} \frac{5\pi}{2} \frac{\Delta^2}{\mu^2} \frac{(\hbar v q)^3}{(\hbar\omega)^3} \frac{\hbar v}{\mu}
\left(\frac{\hbar\omega}{\Delta^2}+\frac{1}{\eta}\frac{2\pi\sqrt{(\hbar\omega)^2-4\Delta^2} \Theta[\hbar\omega-2\Delta]}{\hbar\omega}\right)
\end{equation}
\end{widetext}
where $\eta$ is a small positive parameter, which is set to be $5\times10^{-2}$~meV in the calculation of the main text. Similarly, the different factor of angular integral for $\gamma_x$ and $\gamma_y$ results in $\sigma^{(2)}_{xyy}=\sigma^{(2)}_{yyx}=\sigma^{(2)}_{xxx}/5$.

\end{document}